\documentclass[aps,pre,twocolumn,groupedaddress]{revtex4-1}

\usepackage{graphicx}

\newcommand{\etc}{\textit{etc.}}
 \newcommand{\ie}{\textit{i.e.\ }}
 \newcommand{\eg}{\textit{e.g.\ }}
 
 \newcommand{\vs}{\textit{vs.}}
 
 \newcommand{\Eq}[1]{Eq.\ (\ref{#1})}

 \newcommand{\Fig}[1]{Fig.\ \ref{#1}}

\begin{document}

% Use the \preprint command to place your local institutional report
% number in the upper righthand corner of the title page in preprint mode.
% Multiple \preprint commands are allowed.
% Use the 'preprintnumbers' class option to override journal defaults
% to display numbers if necessary
%\preprint{}

%Title of paper
\title{Percolation thresholds on 3-dimensional lattices with 3 nearest neighbors}

% repeat the \author .. \affiliation  etc. as needed
% \email, \thanks, \homepage, \altaffiliation all apply to the current
% author. Explanatory text should go in the []'s, actual e-mail
% address or url should go in the {}'s for \email and \homepage.
% Please use the appropriate macro for each each type of information

% \affiliation command applies to all authors since the last
% \affiliation command. The \affiliation command should follow the
% other information
% \affiliation can be followed by \email, \homepage, \thanks as well.
\author{Jonathan Tran$^\dagger$, Ted Yoo$^\dagger$, Shane Stahlheber, Alex Small}
\email[]{arsmall@csupomona.edu}

\homepage[]{\\https://sites.google.com/site/physicistatlarge/}

\thanks{$^\dagger$ These authors contributed equally}
%\altaffiliation{}
\affiliation{Department of Physics and Astronomy, California State
Polytechnic University, Pomona, CA 91768}

%Collaboration name if desired (requires use of superscriptaddress
%option in \documentclass). \noaffiliation is required (may also be
%used with the \author command).
%\collaboration can be followed by \email, \homepage, \thanks as well.
%\collaboration{}
%\noaffiliation

\date{\today}

\begin{abstract}
We present a study of site and bond percolation on periodic lattices
with 3 nearest neighbors per site.  We have considered 3 lattices,
with different symmetries, different underlying Bravais lattices,
and different degrees of longer-range connections. As expected, we
find that the site and bond percolation thresholds in all of the
3-connected lattices studied here are significantly higher than in
diamond. Interestingly, thresholds for different lattices are
similar to within a few percent, despite the differences between the
lattices at scales beyond nearest and next-nearest neighbors.
\end{abstract}

% insert suggested PACS numbers in braces on next line
% insert suggested keywords - APS authors don't need to do this
%\keywords{}

%\maketitle must follow title, authors, abstract, \pacs, and \keywords
\maketitle

% body of paper here - Use proper section commands
% References should be done using the \cite, \ref, and \label commands
\section{Introduction}
% Put \label in argument of \section for cross-referencing
%\section{\label{}}

Percolation is one of the simplest phase transitions known: sites on
a lattice are occupied at random until there is a path that can be
traversed from one end of the lattice to the other, traveling only
neighbor-to-neighbor on occupied sites\cite{Stauffer1994}.  The set
of sites along this path or connected to this path (via
neighbor-to-neighbor steps along occupied sites) is called the
spanning cluster. In the limit of a large system size (linear
dimension $L \gg 1$), the probability of forming a spanning cluster
via random occupation of sites goes to zero below a critical
occupation probability per site $p_c$ and unity above $p_c$.
Percolation models have been used to study a great many phenomena,
from transport in porous media to biological systems to
gelation\cite{Sahimi1994}.

A well-known trend in percolation on periodic 3D lattices is that
$p_c$ increases as the coordination number $z$ (number of nearest
neighbors per site) decreases\cite{Galam1996}.  (See
Table~\ref{tab:table1} near the end for examples and more
references.)  This trend arises from the fact that, when there are
fewer neighbors per site, there are fewer ways to navigate around an
obstacle.  Consequently, more sites have to be occupied to ensure a
path spanning from one end of the system to the other. To our
knowledge, the lowest coordination number studied for percolation on
simple 3D lattices is $z=4$, the most prominent example being the
diamond lattice with $p_c\approx 0.43$ for site
percolation\cite{Silverman1990} and $\approx 0.388$ for bond
percolation\cite{Vyssotsky1961}.

However, $z=4$ is not the lowest possible coordination number. The
lowest possible non-trivial coordination number is $z=3$\footnote{In
order to have an average coordination number less than 3, one would
need at least some sites with $z=2$, but a site with $z=2$ is
equivalent to a single bond between two other sites. A physical
analogy would be the role of an oxygen atom in an SiO$_2$ crystal.
So, any lattice with average $z < 3$ must be equivalent to taking a
lattice with $z\geq 3$ and placing sites with $z=2$ along some of
the bonds.}, and this coordination number has been realized in
interesting families of 3D lattices, namely the $(n,3)$ families of
lattices studied by A.\ F.\ Wells\cite{Wells1977}.  The $3$ refers
to the number of nearest neighbors (bonds) per site, and $n$ is the
smallest number of steps that one would have to take along the
lattice sites to return to the same point.  In lattice families with
multiple members, $(n,3)$ is followed by a letter, \eg $(10,3)$-a,
$(10,3)$-b, \etc\ A total of 30 lattices have been identified in
these families, with $n$ values of 7, 8, 9, 10, and 12.  The $(n,3)$
lattices offer a chance to study the interplay between nearest
neighbors, higher-level connectivity (via $n$), and other aspects of
lattice geometry (\eg\ the differences between, say, $(10,3)$-a and
$(10,3)$-b) for the lowest non-trivial $z$ value.  We are only aware
of one study of percolation on lattices in this family, namely the
$(10,3)$-a lattice\cite{Paterson2002}, and that study focused on
invasion percolation and trapping, rather than the standard site and
bond percolation problems.

Recently, the $(10,3)$-a lattice has attracted additional attention
because of its unusually high symmetry, possessing a property known
as ``strong isotropy."\cite{Sunada2008}  Only one other 3D lattice
(diamond), shares this property with $(10,3)$-a.  The $(10,3)$-a
structure is observed in a number of interesting systems, \eg\ block
copolymers\cite{Bates2005}, and it has been proposed as a possible
structure for a metastable phase of carbon\cite{Itoh2009}. Motivated
by this recent attention to $(10,3)$-a and related lattices, as well
as the general question of how high the percolation threshold can be
for a 3D lattice with $z=3$, we have studied 3 lattices in this
family:  $(10,3)$-a, its close relative $(10,3)$-b, and $(8,3)$-a.
These particular lattices span multiple $n$ values, and are
especially easy to study because they can be realized with bonds of
uniform length and $120^{\circ}$ bond angles, making them easy to
construct with ball and stick models.

In what follows, we first discuss the geometries of our three
representative lattices in their simplest forms: equal bond lengths
and $120^{\circ}$ bond angles.  We illustrate deformations of the
lattices that preserve the topology (connections between nearest
neighbors) but enable a mapping onto a cubic lattice, for
convenience in enumerating sites in a calculation.  We then
summarize key features of the Newman-Ziff algorithm\cite{Newman2001}
used to determine the site and bond percolation thresholds of the
lattices, and present the computed percolation thresholds.  Finally,
we compare our results with other lattices.

\section{The lattices under study}

\subsection{The (10, 3)-a lattice}

\begin{figure}[h]
 \includegraphics[scale=0.35]{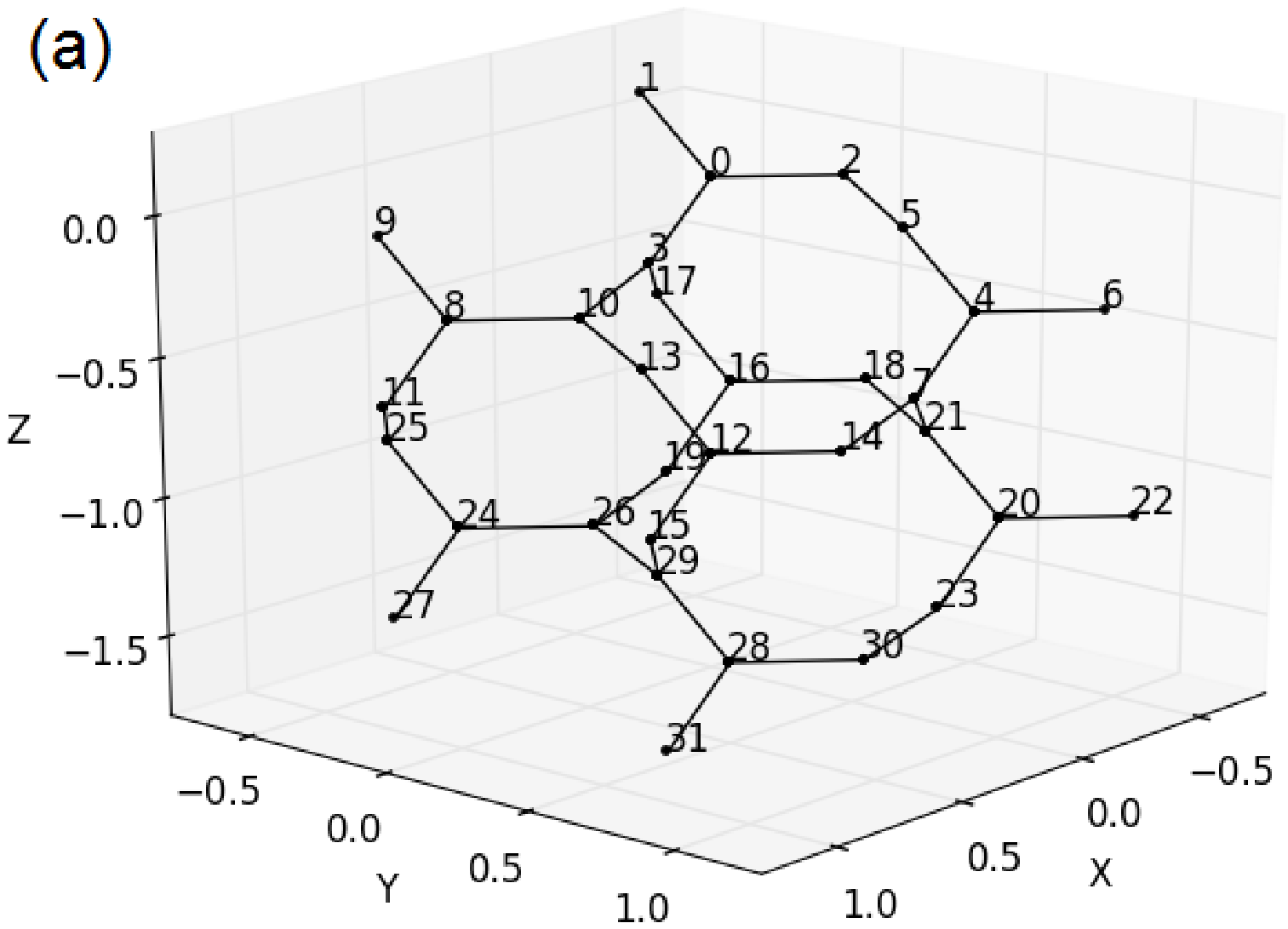}
 \includegraphics[scale=0.35]{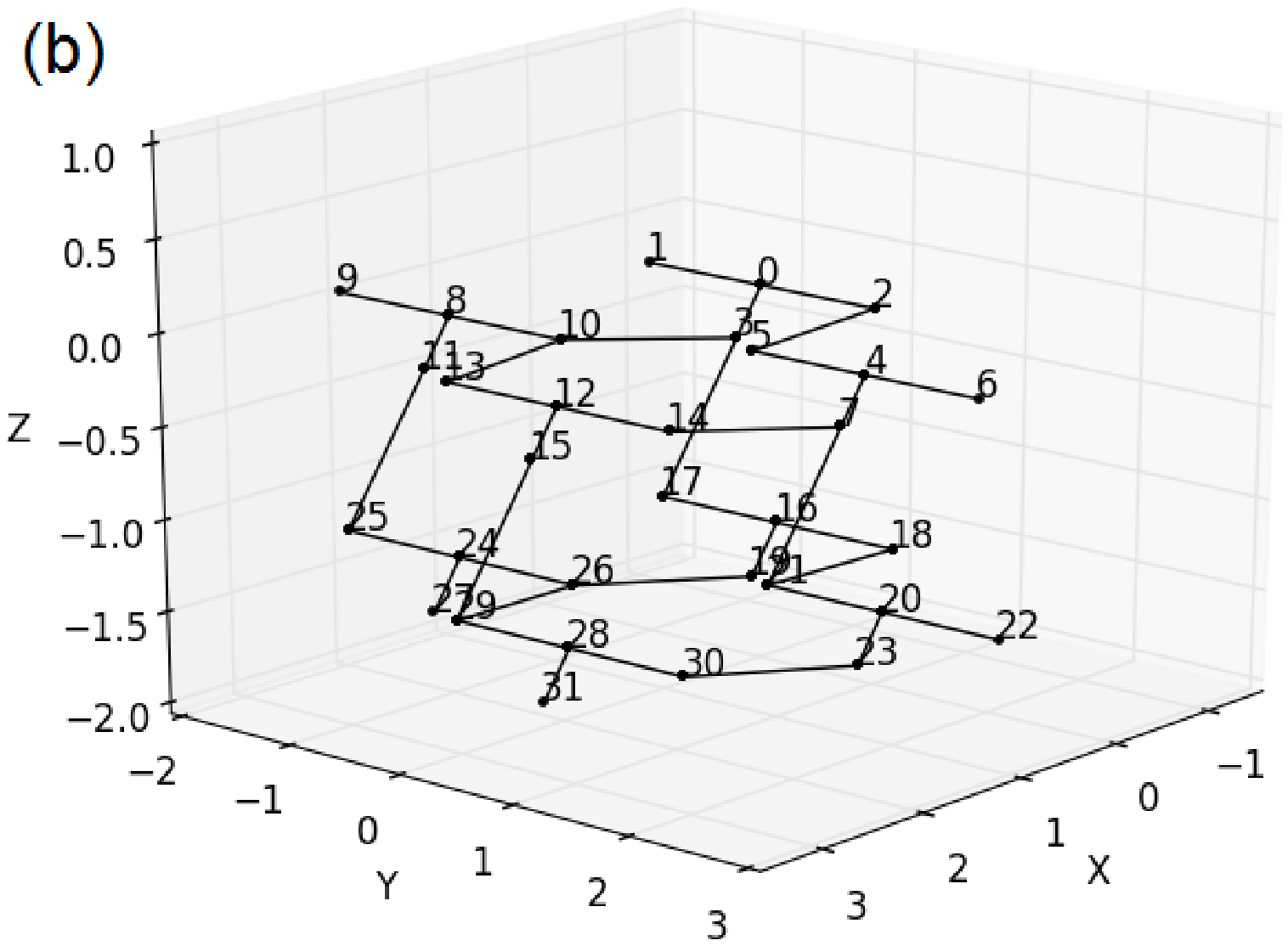}
 \includegraphics[scale=0.35]{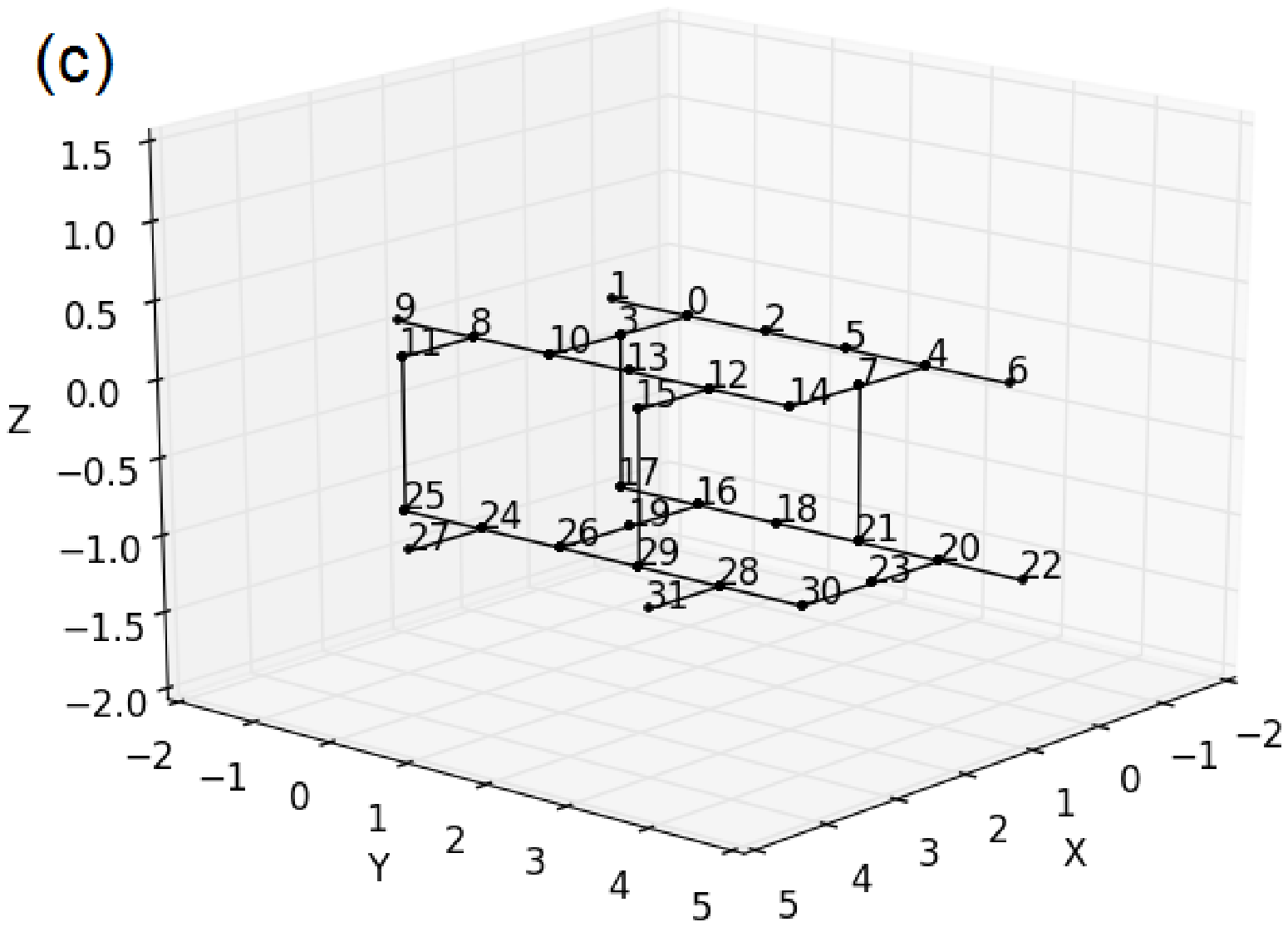}
 \caption{\label{fig:103alattice} (a) The (10,3)-a lattice, with sites numbered. (b) and (c) The deformations used to map the lattice onto a cubic
 grid for computational convenience.}
\end{figure}

The (10, 3)-a lattice (shown in \Fig{fig:103alattice}(a)) can be
thought of as a body-centered cubic (bcc) lattice with a 4-atom
basis.  If we work in a coordinate system where the sites of the bcc
lattice are at the corners of a cube, one site is at the origin
$(0,0,0)$, and the edges are of unit length and parallel to the
standard $x$, $y$, and $z$ axes, then the 3 other atoms in the basis
are at $\left(0, -\frac{1}{4},\frac{1}{4} \right)$, $\left(
-\frac{1}{4},\frac{1}{4}, 0 \right)$, and
$\left(\frac{1}{4},0,-\frac{1}{4} \right)$. The atoms correspond to
sites 0 through 3 in the figure (\ie\ the atoms at the origin and
its 3 nearest neighbors), they sit in the (111) plane, and the bonds
to them are separated by $120^{\circ}$ angles.

In percolation theory, the geometry of the bonds is less important
than the presence of the bonds.  A deformation of the lattice and
bonds that preserves the connections between sites will not change
the percolation threshold of the lattice.  In the Newman-Ziff
algorithm that we used, all of the relevant information on the
lattice is stored in the \texttt{"nn"} array, which contains the
labels of the nearest neighbors bonded to each site in the lattice.
For convenience in enumerating sites, we have deformed the lattice
to fit onto a simple cubic grid, while preserving the bonding
structure of the lattice.  Our numbering system for sites, and the
deformations used to fit them onto a cubic grid, are shown in
\Fig{fig:103alattice}(b) and \Fig{fig:103alattice}(c).

\subsection{The (10,3)-b lattice}

The (10,3)-b lattice is unusual, because even if we specify unit
bond lengths and 120 degree bond angles we still have an
unconstrained structural degree of freedom: Suppose that we define
the positions of all of the lattice sites and also the basis. We can
then deform the lattice in a manner that uniformly changes the
spacing between lattice planes, while also displacing the atoms in
the lattice planes, without changing any bond lengths or bond
angles. We have chosen to describe this lattice in a form that
maximizes the symmetry, as this form lends itself to easier
visualization with ball-and-stick models.  In this form, the lattice
is body-centered tetragonal, with lattice vectors $(\sqrt{3},0,0)$,
$(0,\sqrt{3},0)$, and $\left(\frac{\sqrt{3}}{2},\frac{\sqrt{3}}{2},3
\right)$.

\begin{figure}[h]
 \includegraphics[scale=0.35]{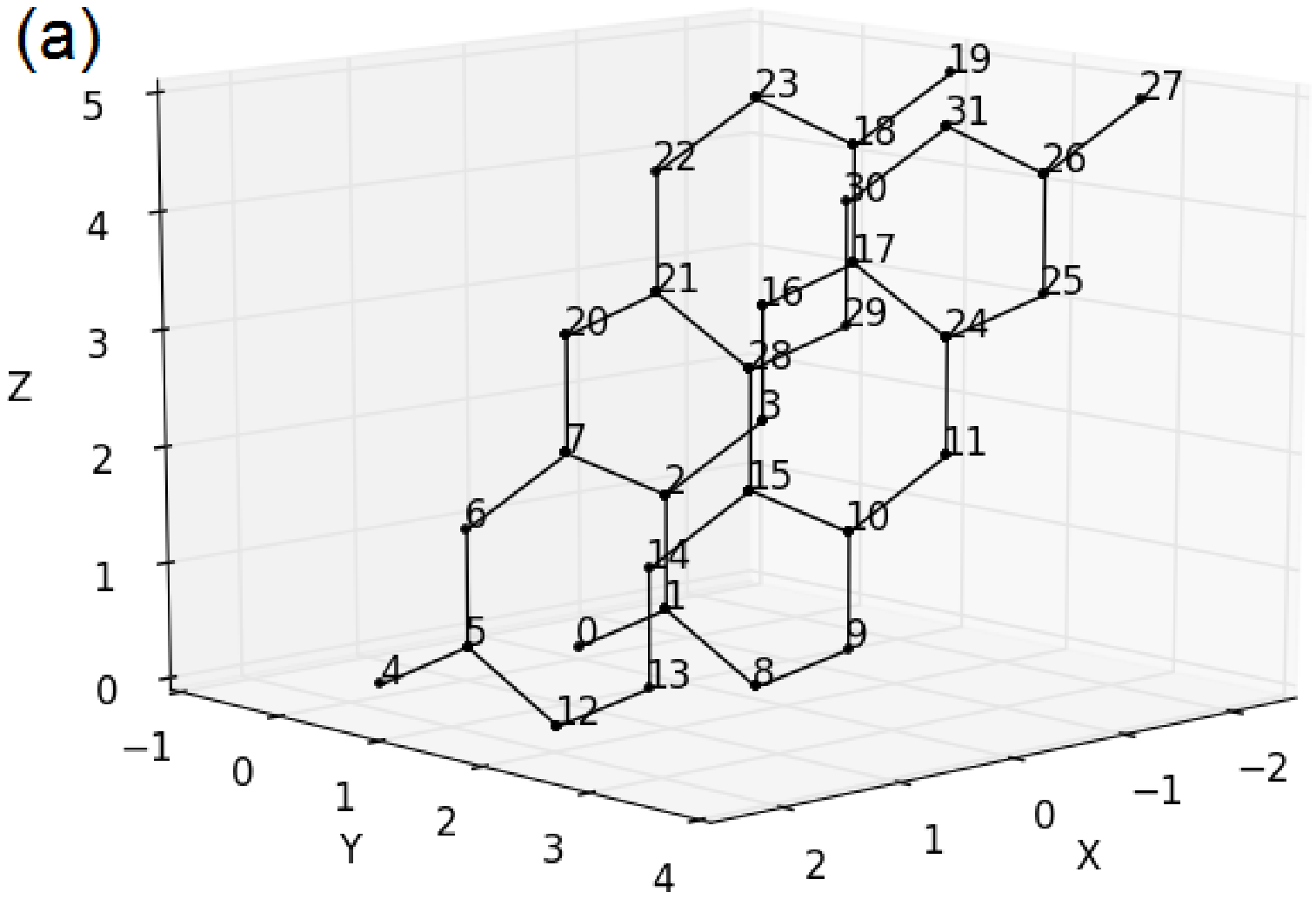}
 \includegraphics[scale=0.35]{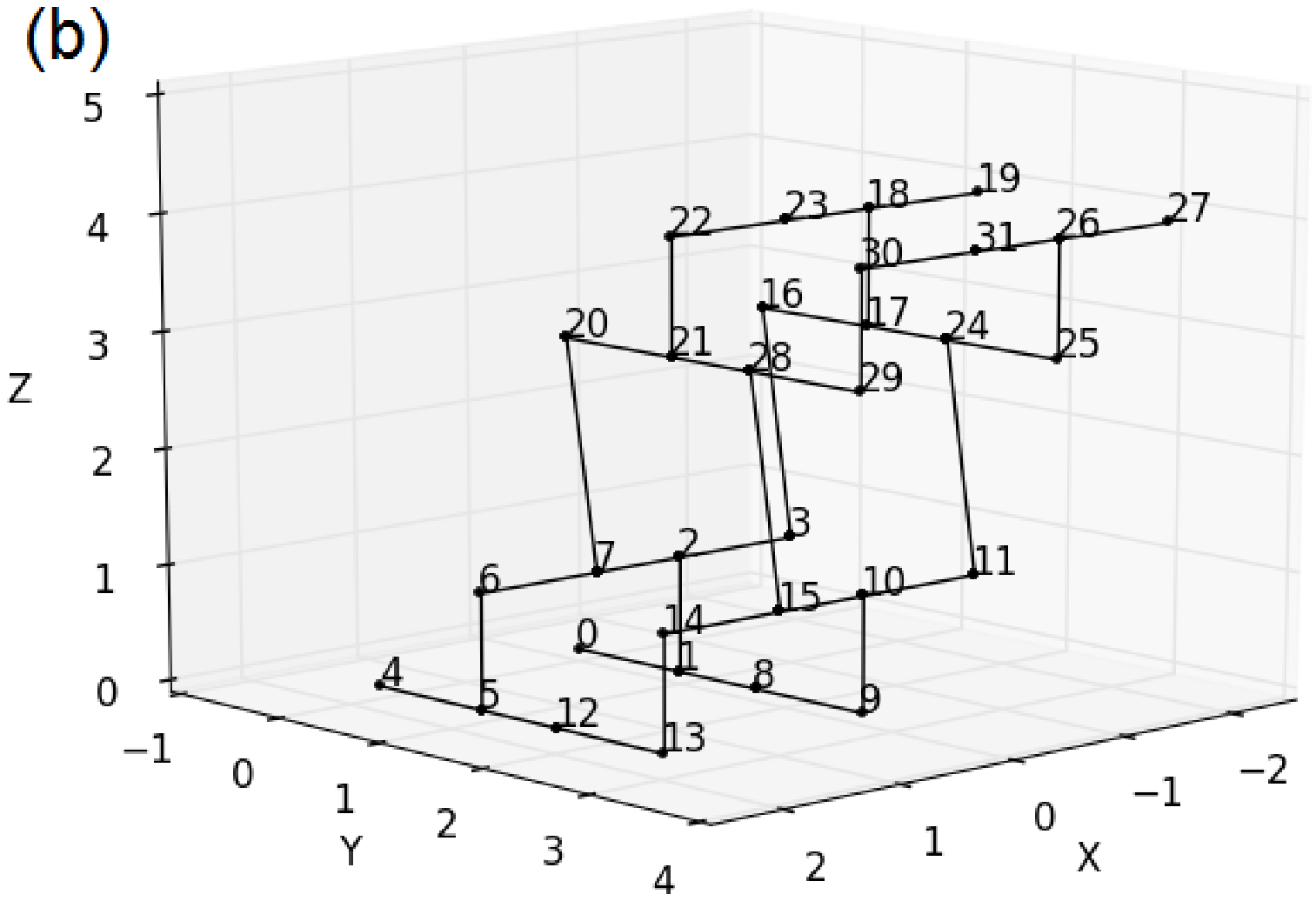}
 \includegraphics[scale=0.35]{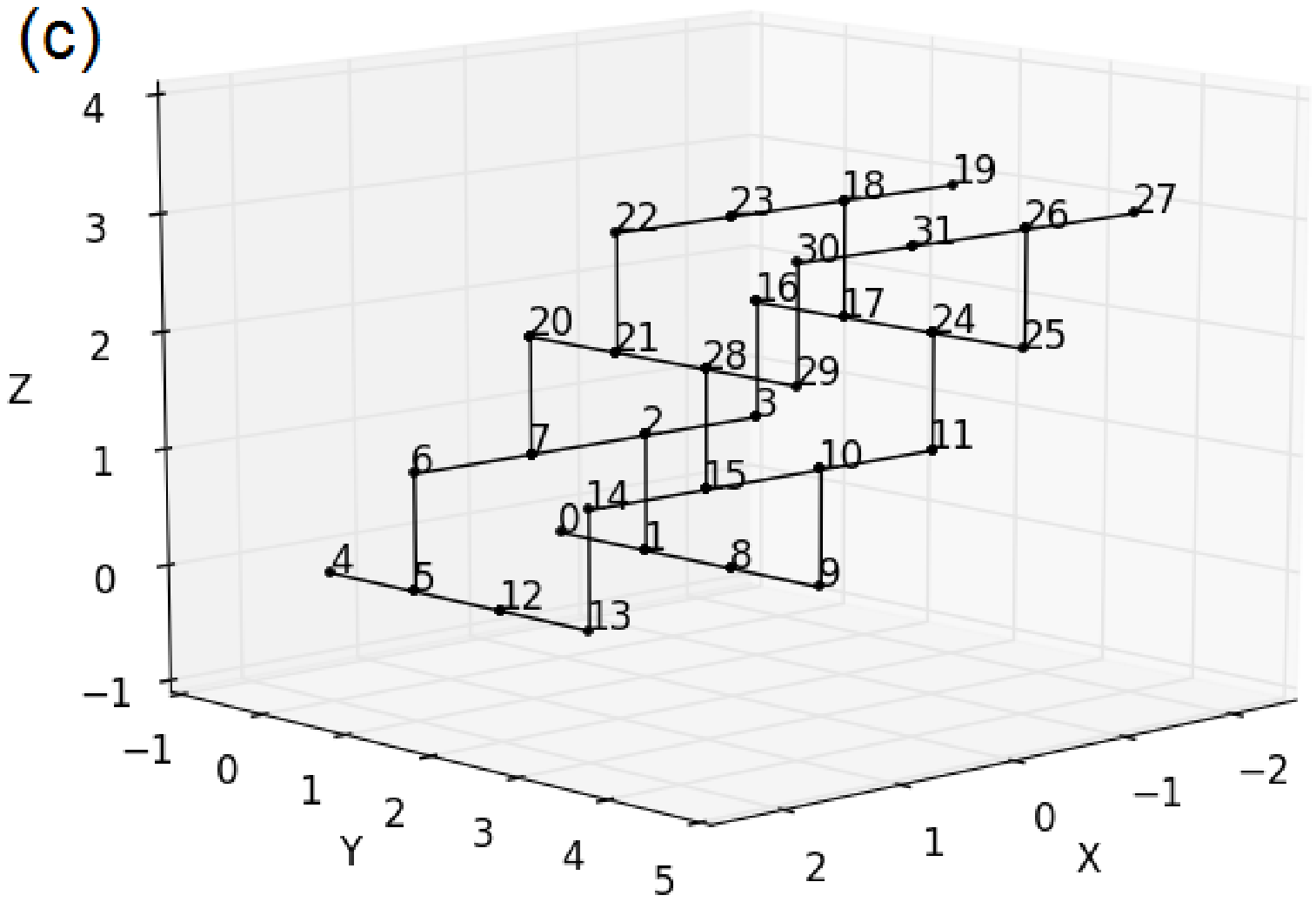}
 \caption{\label{fig:103blattice} (a) The (10,3)-b lattice, with sites numbered. (b) and (c) The deformations used to map the lattice onto a cubic
 grid for computational convenience  }
\end{figure}

The 4-atom basis is more complicated. Instead of a central atom with
3 neighbors, the basis is a chain of 4 atoms, located at $(0,0,0)$,
$\left( 0,\frac{\sqrt{3}}{2}, \frac{1}{2} \right)$, $\left(
0,\frac{\sqrt{3}}{2}, \frac{3}{2} \right)$, and $\left(
-\frac{\sqrt{3}}{2}, \frac{\sqrt{3}}{2}, 2 \right)$.  These
correspond to sites 0 through 3 in the figure. As before, we also
deform this lattice to represent it on a simple cubic grid. The
numbering system and deformation steps are illustrated in
\Fig{fig:103blattice}(b) and (c).

\subsection{The (8,3)-a lattice}

\begin{figure}[h]
 \includegraphics[scale=0.35]{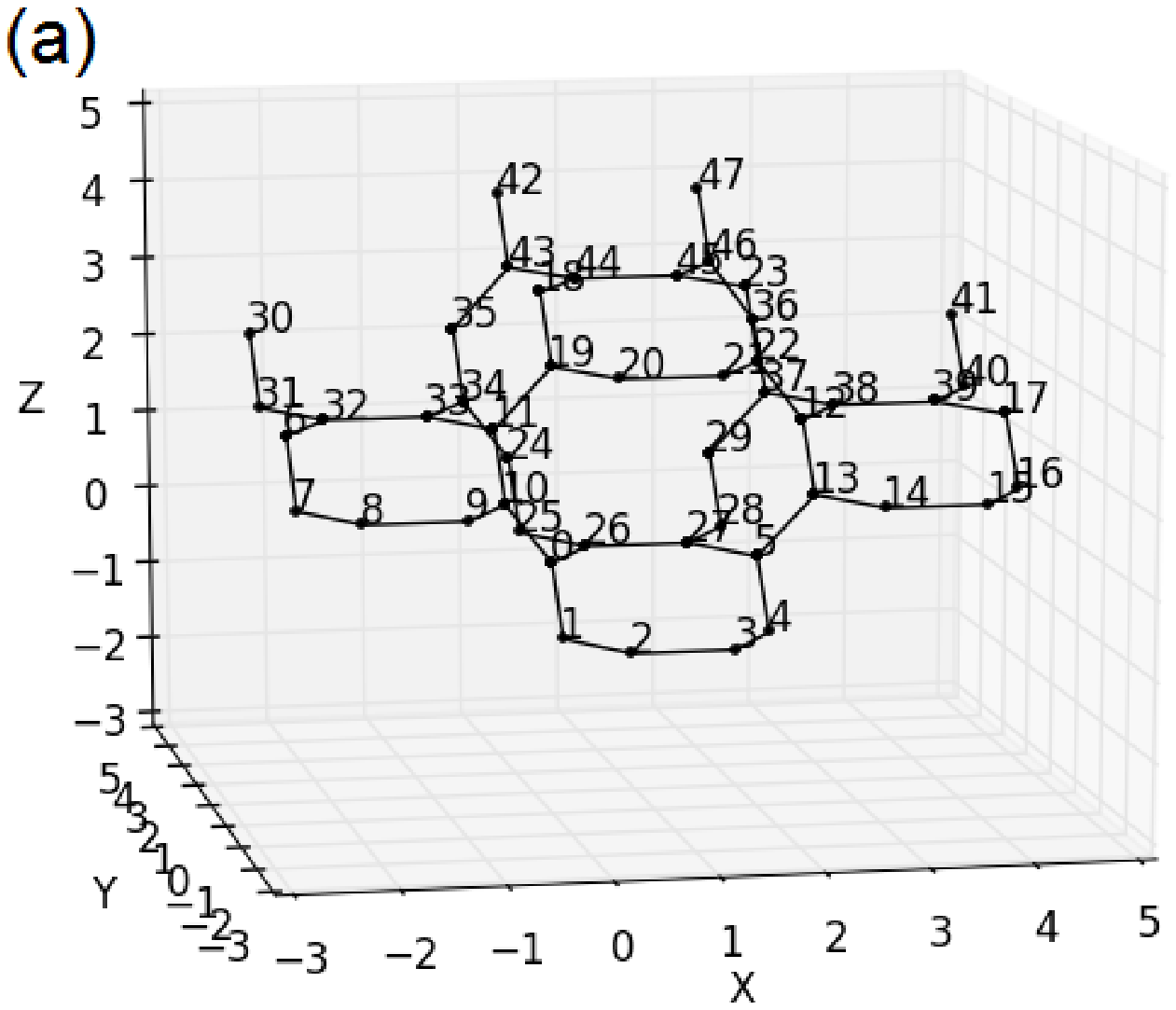}
 \includegraphics[scale=0.35]{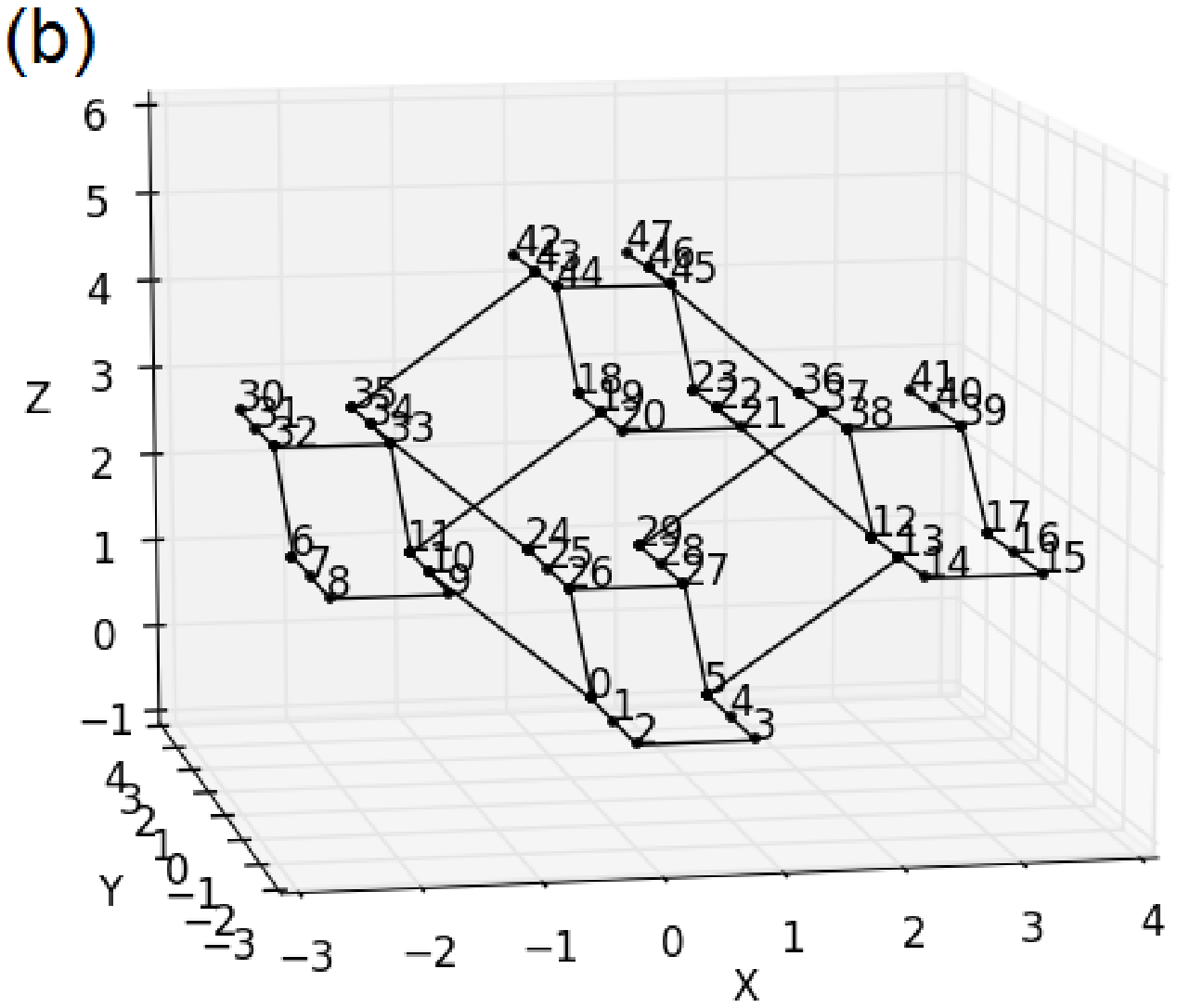}
 \includegraphics[scale=0.35]{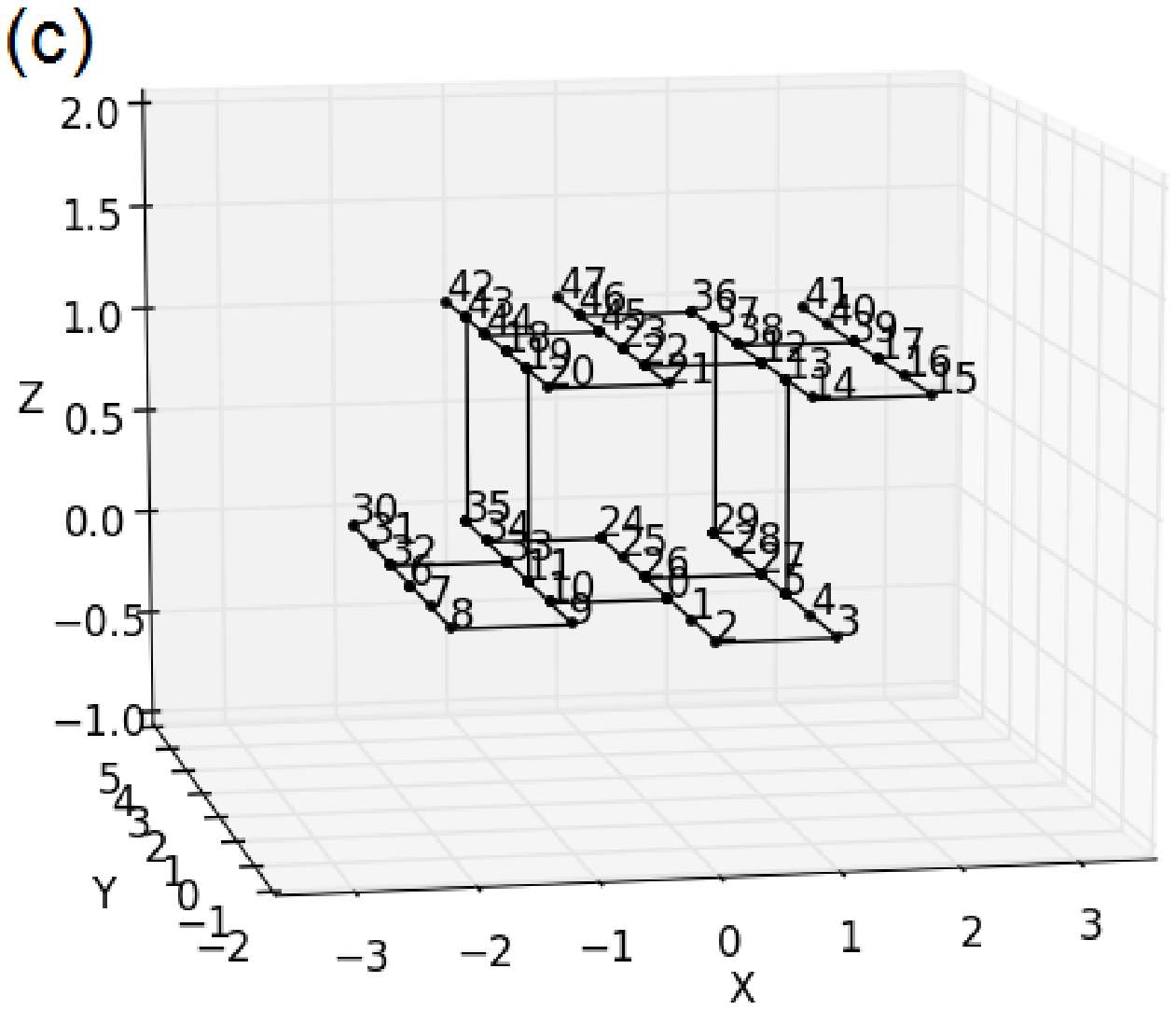}
 \caption{\label{fig:83alattice} (a) The (8,3)-a lattice, with sites numbered. (b) and (c) The deformations used to map the lattice onto a cubic
 grid for computational convenience  }
\end{figure}

We realized the (8,3)-a lattice, shown in \Fig{fig:83alattice}(a),
in a structure that has hexagonal symmetry, with lattice vectors
$(-5/2, \sqrt{3}/6,2\sqrt{2/3})$, $(5/2, \sqrt{3}/6, 2\sqrt{2/3})$,
and $(0, 4\sqrt{3}/3, \sqrt{2/3})$.  The $x$ axis is the axis of
hexagonal symmetry.

The (8, 3)-a lattice has a six-atom basis, with unit bond lengths
and atoms located at $(0,0,0)$, $(0,-\sqrt{3}/3, -\sqrt{2/3})$,
$(1/2,-5\sqrt{3}/6,-\sqrt{2/3})$, $(3/2,-5\sqrt{3}/6,-\sqrt{2/3})$,
$(2,-\sqrt{3}/3, -\sqrt{2/3})$, and $(2,0,0)$.  These correspond to
sites 0 through 5 in the figure. The steps used to deform the
(8,3)-a lattice so that it fits onto a cubic grid are illustrated in
\Fig{fig:83alattice}(b) and (c).

%\subsection{The (12,3) lattice}
%
%The (12,3) lattice, illustrated in \Fig{fig:123lattice}, is similar
%to (10,3)-b, in that it also has a degree of freedom:  By rotating
%the lattice planes, one can vary their spacing without changing any
%bond angles or bond lengths.
%
%The steps used to deform the (12,3) lattice so that it fits onto a
%cubic grid are illustrated in \Fig{fig:123deformation}

\section{Simulations}

\subsection{Description of the algorithm}

We used the Newman-Ziff algorithm for identifying the percolation
threshold of finite-sized clusters\cite{Newman2001}.  In brief, the
algorithm works by occupying sites (or bonds) on a lattice of $N$
one-at-a-time in a random order.  Relationships between sites are
defined in the array \texttt{nn}, which indicates the site numbers
for the nearest neighbors of each site.  At each step, after a site
is occupied we check whether the occupation of the $n$th site
produces a cluster that wraps around the entire system. Each cluster
is assigned a ``pointer'' to a root (or parent) site, corresponding
to the first site in the cluster, and also a vector that points in
the direction of the parent site. As clusters grow and merge,
pointers and vectors are updated to the root of the largest cluster
involved in the merger. Wrapping is detected when a newly-occupied
site (1) joins two portions of the same cluster and (2) the vectors
going from the joined portions of the cluster to the root differ by
something other than the displacement vector between the sites.  We
consider wrapping between parallel faces of the computational volume
(\ie\ along the $x$, $y$, and $z$ directions) as well as diagonal
wrapping. (See the paper by Newman and Ziff\cite{Newman2001} for
more details.) The iteration ends when wrapping is detected. Another
lattice is then generated, and the process is repeated, until $N_L$
lattices (typically $10^3$ in our work) have been generated.  Bond
percolation is handled in a completely analogous manner,
substituting bonds for sites.  Two bonds are considered neighbors if
they touch the same site.

This process generates a plot of wrapping probability $R_L$ \vs\
occupation fraction $\phi=n/N$, where $L$ refers to the linear
dimension of the lattice.  In order to determine $R_L$ as a function
of occupation \textit{probability} (the usual quantity of interest
in percolation theory), it is necessary to convolve $R_L$ with a
binomial distribution:
\begin{eqnarray}
    R_L(p) = \sum_{n=0}^N \left( \begin{array}{c} N \\ n \end{array}
    \right) p^n (1-p)^{N-n} R_L(n)
    \label{eq:binomial}
\end{eqnarray}
\Eq{eq:binomial} amounts to a weighted sum over all possible
realizations of an $N$-site lattice with site occupation probability
$p$.  The number of realizations that wrap with $n$ occupied sites
enters the sum via $R_L(n)$.  Different occupation numbers $n$ will
have different likelihoods of being realized at the same occupation
probability $p$, \ie\ different degeneracies, and this enters the
sum via the binomial distributions.  Occupation numbers that are not
close to $N\cdot p$ are unlikely, and hence get low weight from the
binomial distribution.

To implement the algorithm we closely followed the code example
given in the paper by Newman and Ziff, implementing it in Python 2.7
on a quad-core Intel processor in Linux (Ubuntu 12.04). We checked
our code by determining the site and bond percolation thresholds of
the 2D square, 2D honeycomb, and 3D simple cubic lattices.  Adapting
the code to the lattices under study only required modification to
the function that identifies the nearest neighbors of each site, as
well as the vectors from each site to its neighbors.  To validate
the output for the lattices under study, we generated lattices with
small numbers of sites ($\approx 30$) and had the code output the
list of occupied sites when wrapping occurred. We verified by hand
that (1) there was a cluster that wrapped the lattice, and (2) the
removal of the most recently occupied site would cause the cluster
to not wrap.

\begin{figure}[h]
    \includegraphics[scale=0.35]{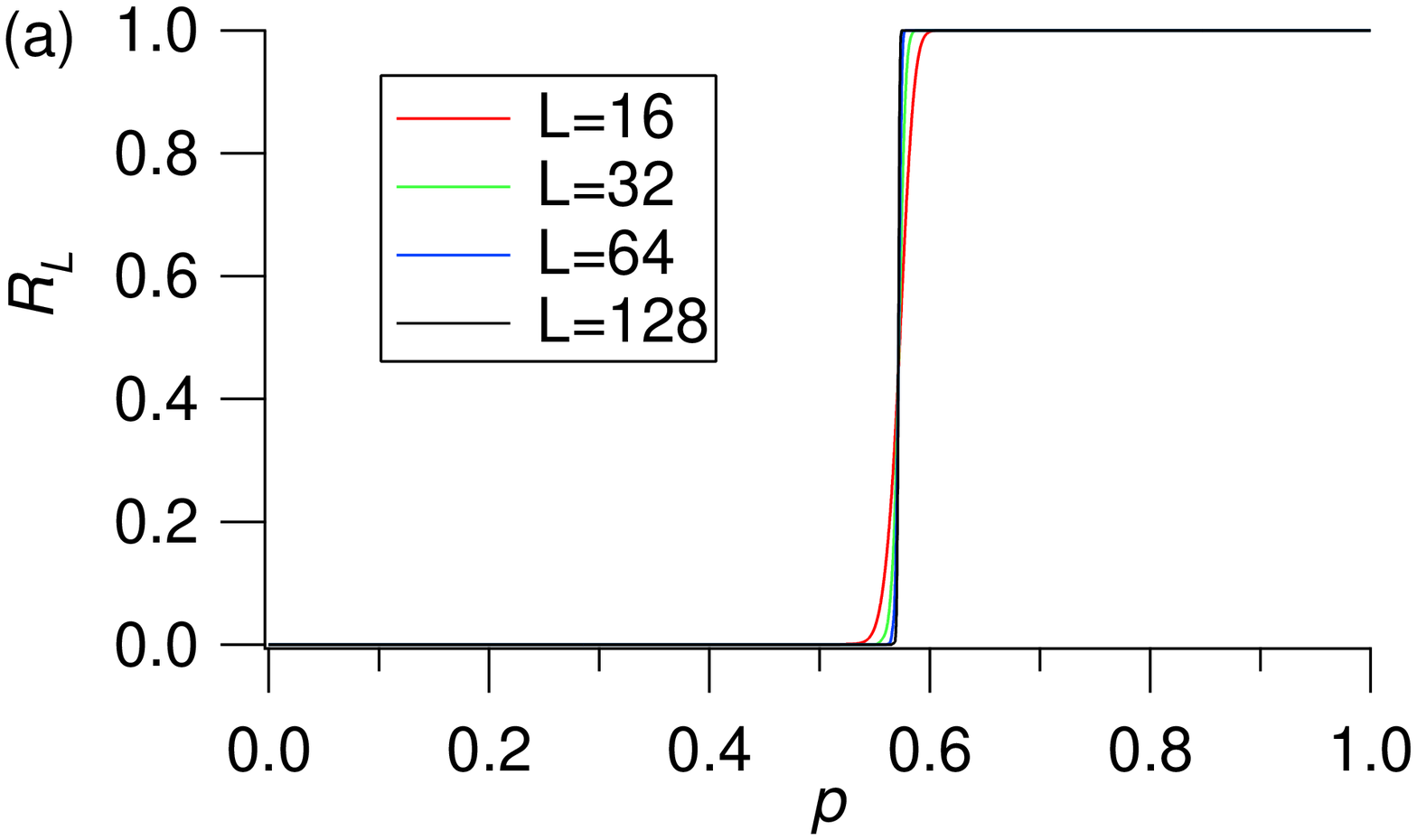}
    \includegraphics[scale=0.33]{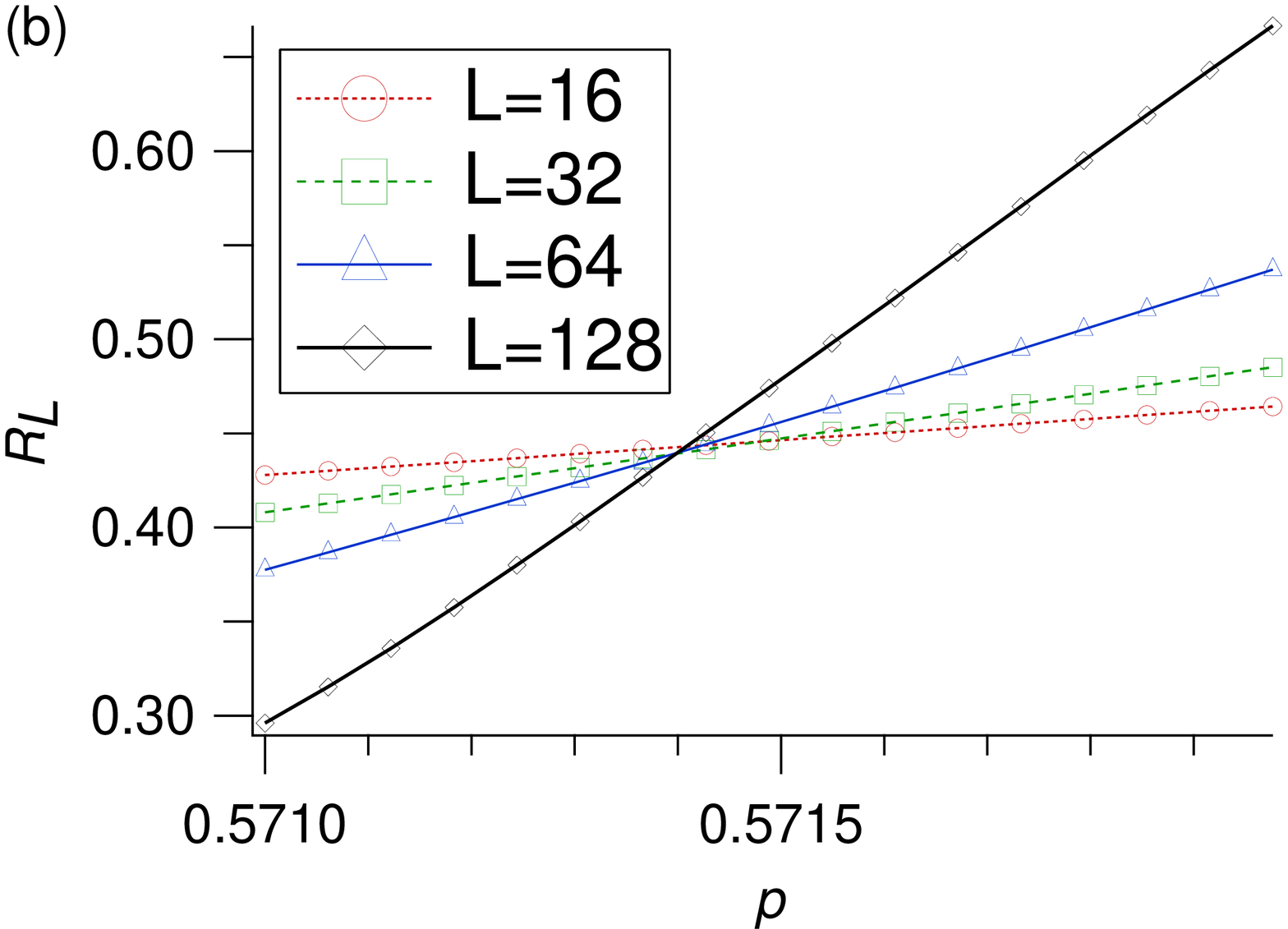}
    \caption{\label{fig:103asitewrapping} (a) Wrapping probability $R_L$ \vs\ occupation
    probability $p$ for sites on the (10,3)-a lattice.  The curves for different system sizes $L$
    cross around $p_c\approx 0.5714$.  (b) Zoom of previous plot.}
\end{figure}

Given a plot of the wrapping probability $R_L(p)$ for different
lattice sizes $L$, it is possible to obtain an estimate of the
percolation threshold $p_c$ by looking for the point where the plot
crosses over from low wrapping probability to high (\eg\ the
steepest point on the plot).  An example for the (10,3)-a lattice is
shown in \Fig{fig:103asitewrapping}. For sufficiently large system
sizes, this estimate of $p_c$ can be quite accurate. More efficient
approaches can obtain high precision and accuracy by comparing
$R_L(p)$ plots for several different system sizes $L$. One common
way of comparing plots at different sizes is to use the scaling
relation $|p_c(L)-p_c|\propto L^{-\omega-\nu}$, where $p_c(L)$ is
obtained from the cross-over of an $R_L(p)$ plot and $\omega$ and
$\nu$ are universal exponents that depend only on
dimension\cite{Ziff2002}.

However, a simpler approach, one that enables a very intuitive
estimate of $p_c$ and the associated uncertainty, is to make plots
with $R_L(p)$ (for fixed $p$) on the vertical axis and $L$ on the
horizontal axis. For $p>p_c$, $R_L(p)$ is an increasing function of
$L$, and for $p<p_c$ $R_L(p)$ is a decreasing function of $L$.  This
follows directly from the fact that $R_L(p)$ becomes more
``step-like'' as $L$ increases.  At $p=p_c$, $R_L(p)$ is independent
of $L$\cite{Martins2003}.  The value of $p_c$ can thus be estimated
simply by looking at the different plots and identifying the
flattest one.  When the plot of $R_L$ oscillated somewhat as a
function of $L$ (due to randomness in the simulations), we
identified the value of $p$ for which $R_L$ oscillates without a
pronounced upward or downward trend.

\subsection{The uncertainty in the percolation threshold}

There is an easy way to determine the uncertainty in $p_c$ based
only on the simulation output, and without any assumptions about
critical exponents. In a plot of $R_L(p)$, the percolation
transition is identified by looking for a value of $p$ at which the
wrapping probability is approximately stationary as the system size
changes. However, the wrapping probabilities are determined from
finite samples in Monte Carlo simulations, and (as described below)
this gives rise to statistical fluctuations in $R_L$ for all $p$
values. A fluctuation $\delta R_L$ shifts the curve along the
horizontal axis by an amount $\delta R_L/(dR_L/dp)$, where $dR_L/dp$
is the slope of the $R_L$ \vs\ $p$ curve.

In order to get the uncertainty in $R_L(p)$, let us first consider
$R_L$ as a function of occupation number $n$ rather than occupation
probability $p$. We use generate $N_L$ lattices with $n$ occupied
sites, and use the Newman-Ziff algorithm to determine the number
$N_L*R_L(n)$ that wrap. If we were to do this process repeatedly, we
would find that the number wrapping obeys a binomial distribution
with mean $N_L R_L(n)$ and variance $N_L R_L(n)(1-R_L(n))$, and so
the fraction $R_L(n)$ that wrap has mean $R_L(n)$ and variance
$R_L(n)(1-R_L(n))/N_L$.

When we go from $R_L(n)$ to $R_L(p)$, the convolution with a
binomial distribution means that $R_L(p)$ is a weighted sum of
different $R_L(n)$ values.  However, $R_L(n+1)$ is not statistically
independent of $R_L(n)$, since it is generated by occupying one more
site (or bond) in each of the lattices used to determine $R_L(n)$.
Near $p_c$, $R_L(n\neq N p_c)$ is approximately a linear function of
$n-Np_c$:
\begin{equation}
    R_L(n)  \approx R_L(Np_c)  + c*(n-Np_c) + \eta(n)
    \label{eq:RL}
\end{equation}
where $c$ is an unknown constant and $\eta$ is a noise term.

This expression for $R_L(n)$ is convolved with a binomial
probability distribution that is approximately Gaussian for large
$N$ (and hence even about $p_c$). The normalization of the
probability distribution means that the convolution with the first
term gives $R_L(Np_c)$.  The symmetry of the peak of the
distribution means that the convolutions with the second and third
terms vanish. We thus conclude that $R_L(p=p_c) \approx
R_L(n=Np_c)$, and so we will use $\sqrt{\frac{R_L(1-R_L)}{N_L}}$ for
the uncertainty in $R_L(p_c)$.  This, in turn gives the uncertainty
in $p_c$ as:
\begin{equation}
    \delta p_c = \sqrt{\frac{R_L(1-R_L)}{N_L}}\left/\frac{dR_L}{dp}
    \right.
    \label{eq:uncert}
\end{equation}
We have neglected the variance of the weighted sum of the noise
terms $\eta(n)$ in \Eq{eq:RL}, but that variance is small because
$\eta(n)$ exhibits correlations for nearby $n$ values:  The fact
that $R_L(n)$ is an increasing function of $n$ means that downward
fluctuations of $R_L(n)$ are limited in magnitude by the
fluctuations of $R_L(n-1)$. The fact that $R_L\leq 1$ reduces the
probability of successive upward fluctuations.

In \Eq{eq:uncert}, we use the slope of the steepest $R_L$ \vs\ $p$
curve.  If the steepest curve were perfectly vertical, fluctuations
of the other curves would be completely irrelevant, and the point of
intersection would remain on that curve at the value of $p$ where it
rises.  Consequently, the finite slope of the steepest curve is the
limiting factor in our determination of $p_c$.

\section{Results}

\begin{figure}[h]
    \includegraphics[scale=0.35]{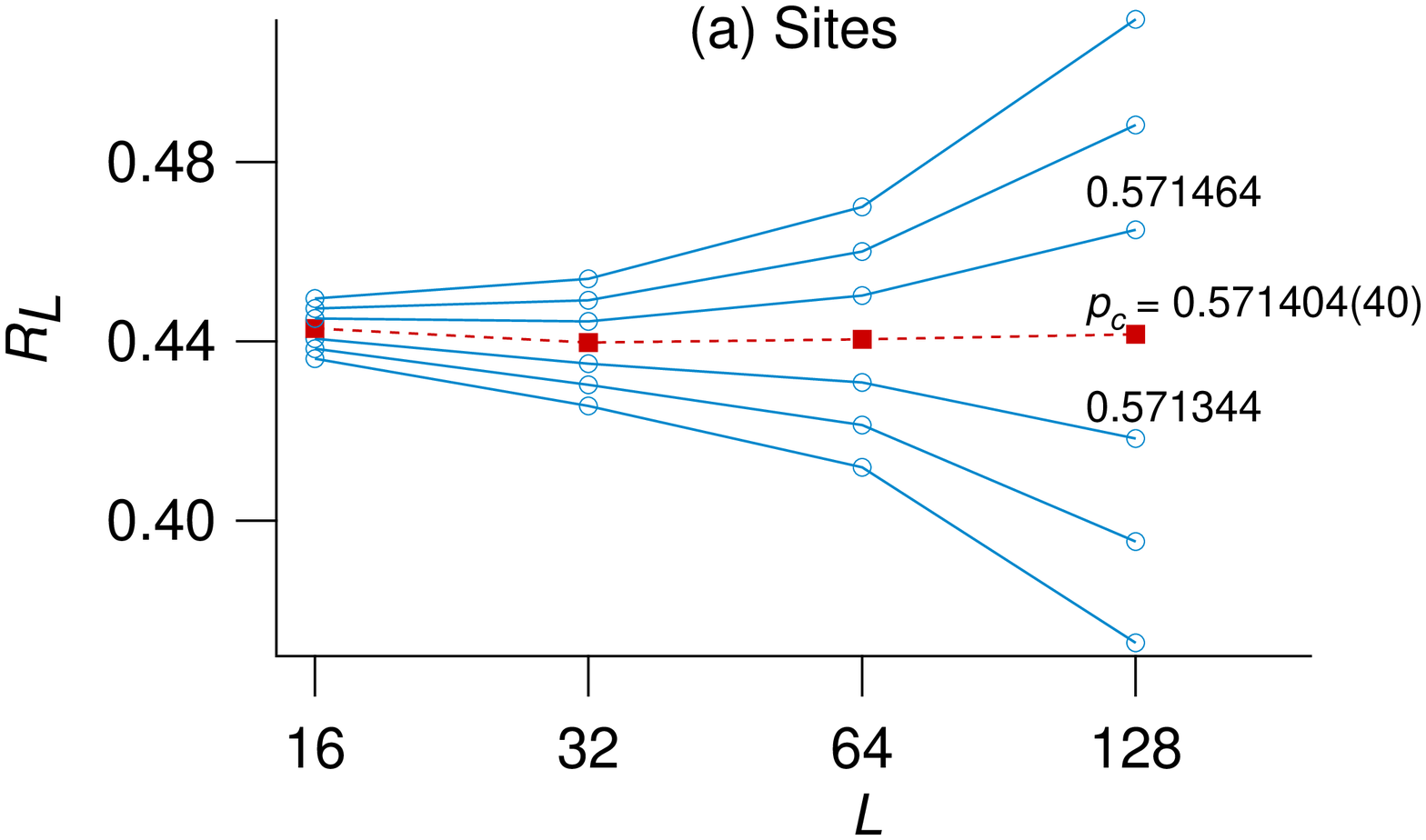}
    \includegraphics[scale=0.35]{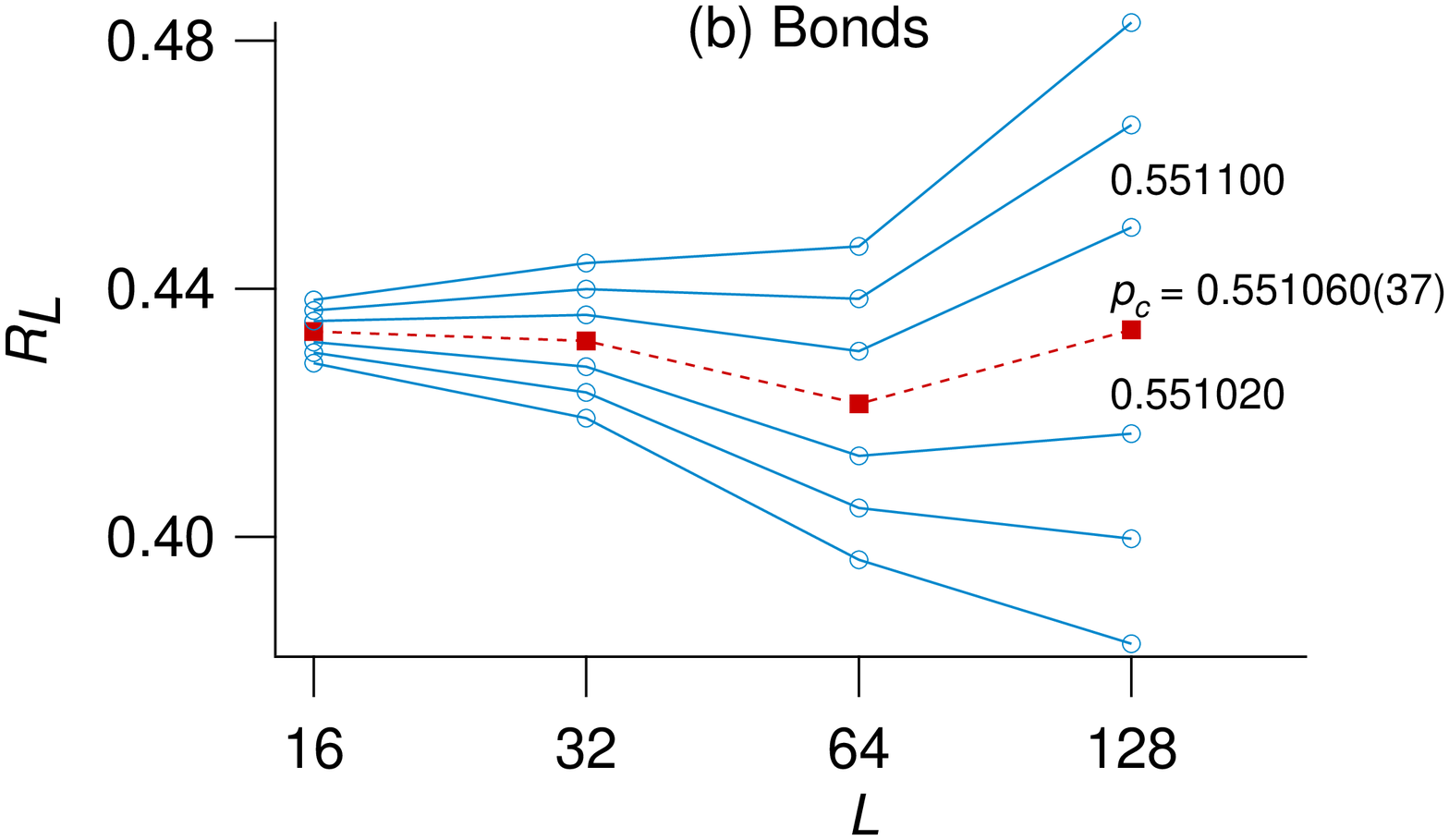}
    \caption{\label{fig:103avsL} Wrapping probability $R_L$
    \vs\ $L$ for (a) site and (b) bond percolation on (10,3)-a lattices, for different occupation probabilities
    $p$.  (Selected $p$ values shown to right of plots.)
     The percolation threshold $p_c$ is the value of $p$ that gives
     the flattest overall trend. The uncertainty
     is determined via \Eq{eq:uncert}.}
\end{figure}

\begin{figure}[h]
    \includegraphics[scale=0.35]{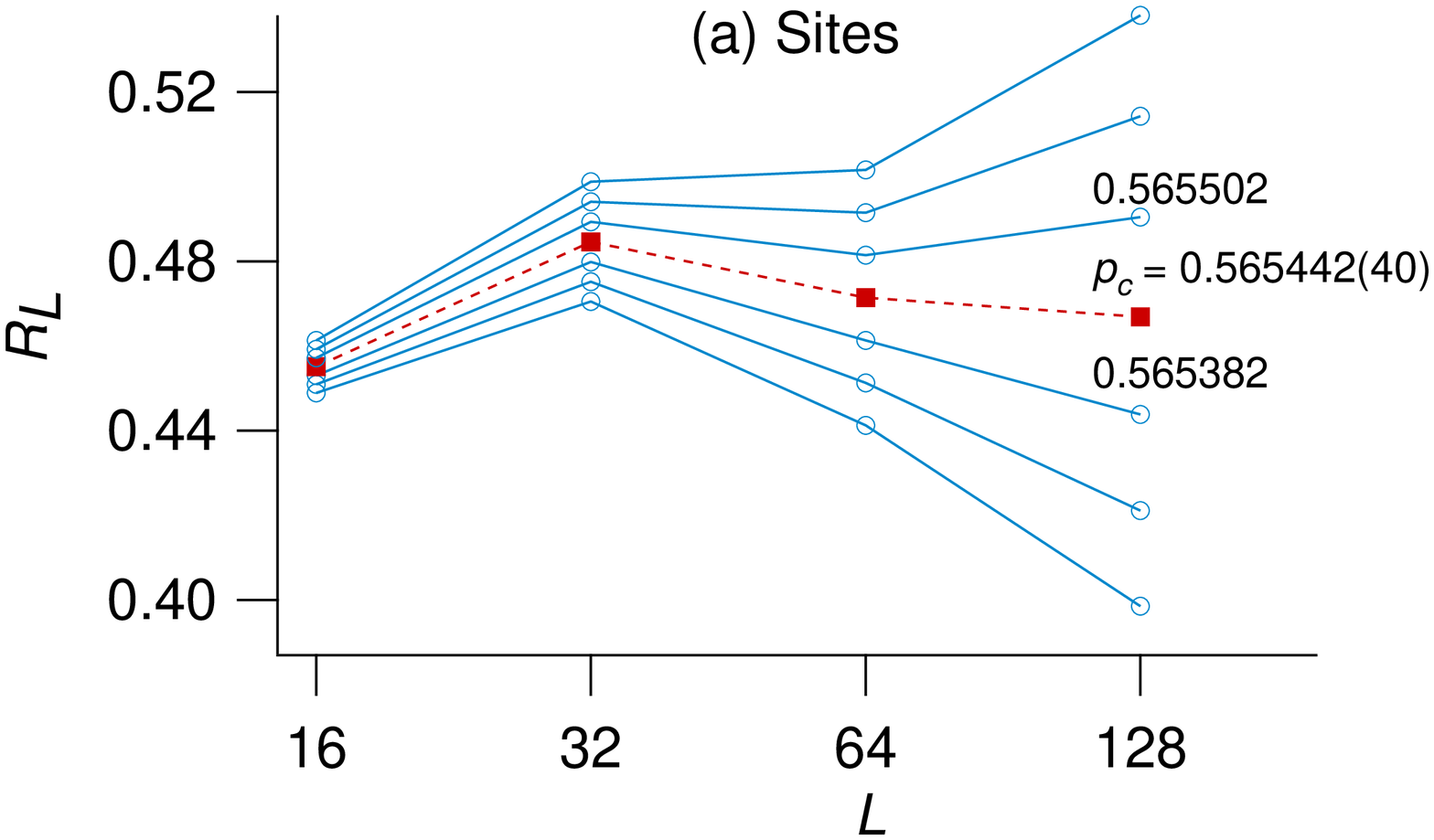}
    \includegraphics[scale=0.35]{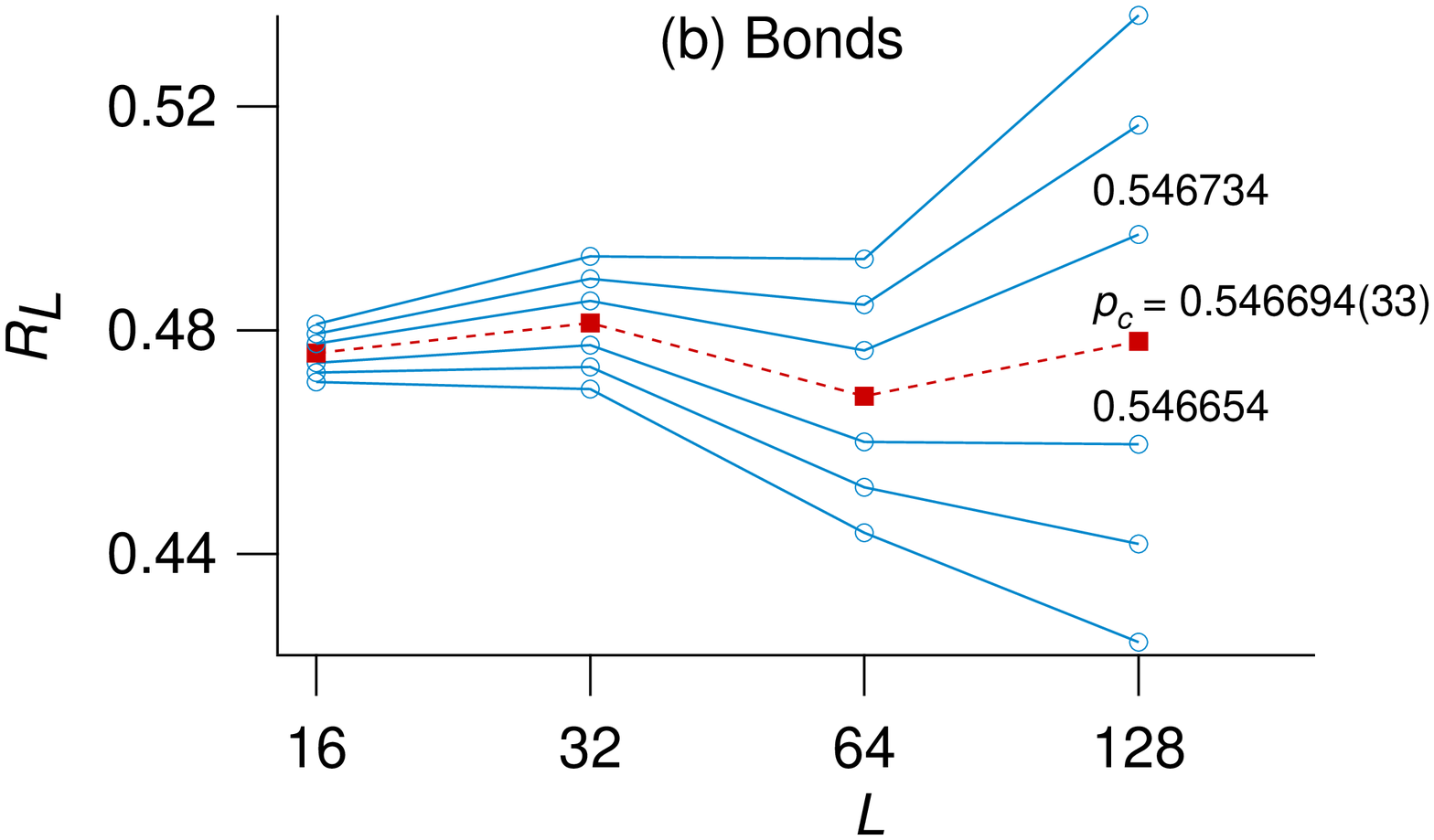}
    \caption{\label{fig:10bvsL} Wrapping probability $R_L$
    \vs\ $L$ for (a) site and (b) bond percolation on (10,3)-b lattices, for different occupation probabilities
    $p$.  (Selected $p$ values shown to right of plots.)
     The percolation threshold $p_c$ is the value of $p$ that gives
     the flattest overall trend. The uncertainty
     is determined via \Eq{eq:uncert}.}
\end{figure}

\begin{figure}[h]
    \includegraphics[scale=0.35]{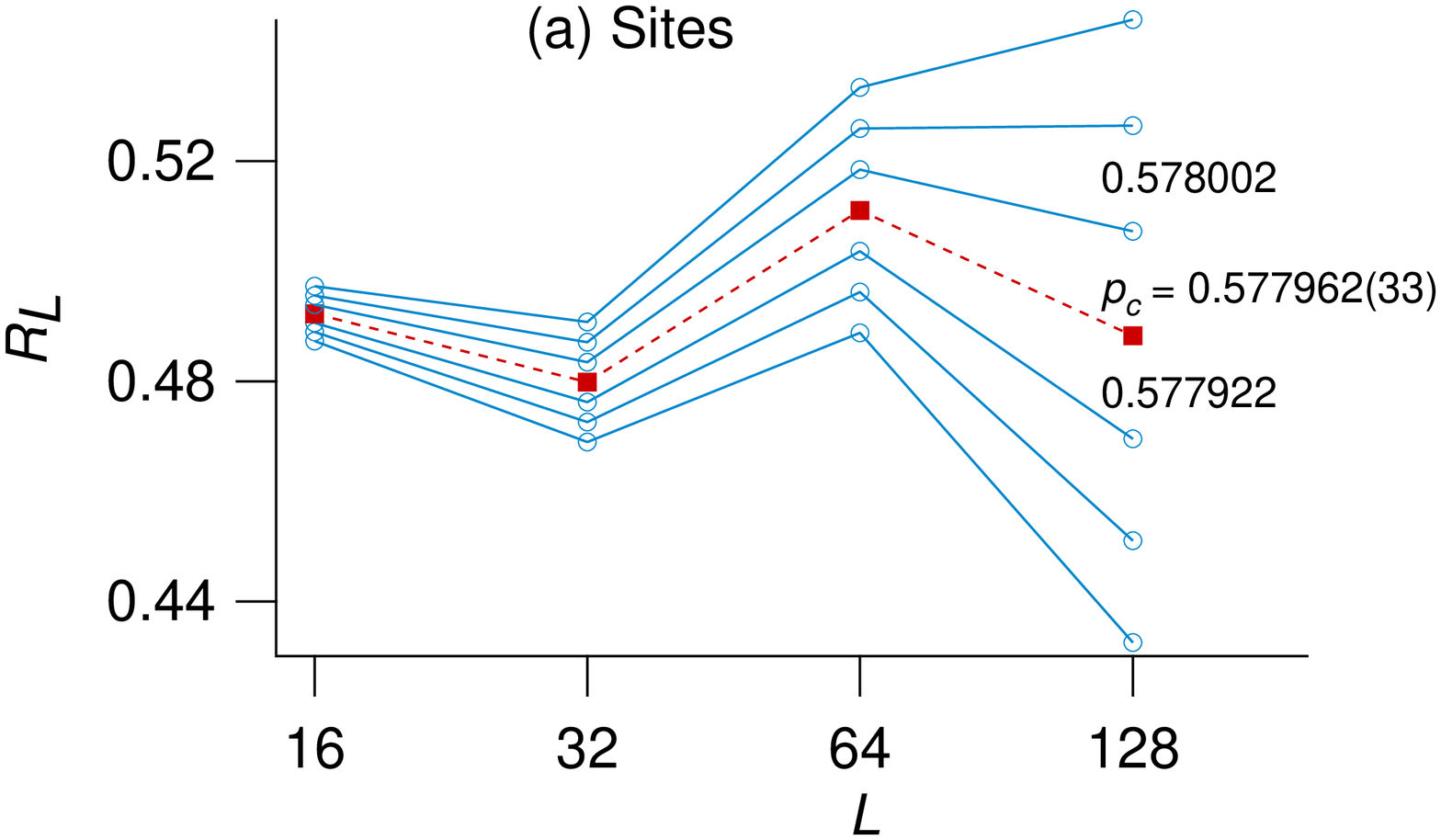}
    \includegraphics[scale=0.35]{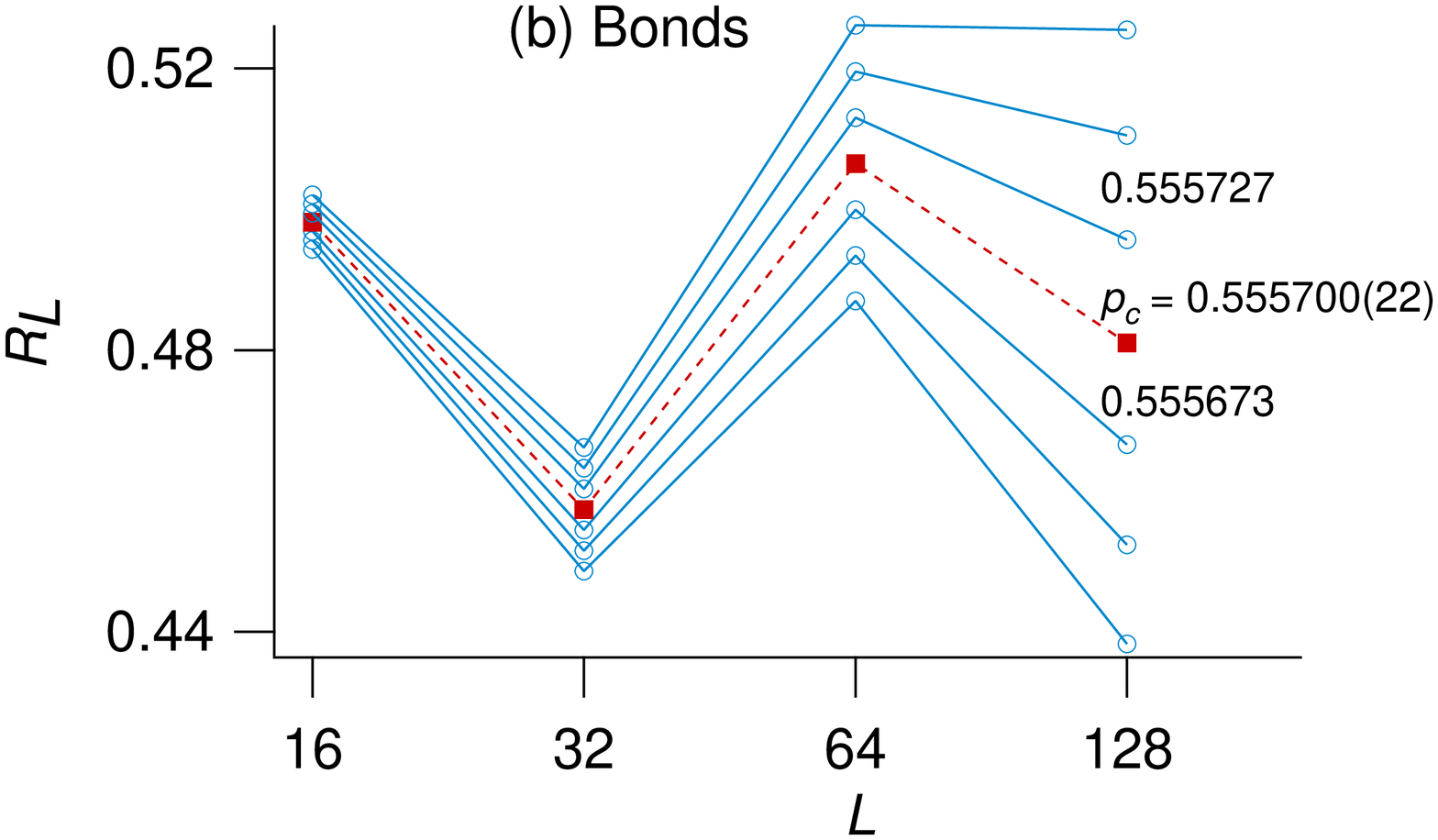}
    \caption{\label{fig:83avsL} Wrapping probability $R_L$
    \vs\ $L$ for (a) site and (b) bond percolation on (8,3)-a lattices, for different occupation probabilities
    $p$.  (Selected $p$ values shown to right of plots.)
     Although there is some oscillation, the percolation threshold $p_c$ is the value of $p$ that gives
     the flattest overall trend. The uncertainty
     is determined via \Eq{eq:uncert}.}
\end{figure}

Figures \ref{fig:103avsL} through \ref{fig:83avsL} show $R_L$ \vs\
$L$ (for different $p$ values), for site and bond percolation on the
lattices under study. The number of unit cells in a realization of
each lattice is $L^3$, so that the number of lattice sites is $4L^3$
(for the (10,3)-a and (10,3)-b lattices) or $6L^3$ (for (8,3)-a).
The number of bonds is $6L^3$ (for (10,3)-a and (10,3)-b) or $9L^3$
(for (8,3)-a). In all plots, site occupation probability $p$ was
incremented in steps of $1/\text{number of sites or bonds for
smallest $L$ value}$.

\begin{table}[h]
\caption{\label{tab:table1} Site and bond percolation thresholds for
important 3-dimensional lattices with different coordination numbers
$z$.  Uncertainties are given in parentheses, and refer to the last
one or two digits, depending on the number of digits in the
uncertainty.  Bibliographic references are in brackets [].
\footnote{For a more exhaustive table of percolation thresholds in
different systems, Prof.\ Robert Ziff regularly updates the
Wikipedia page ``Percolation Threshold." A snapshot from the date of
this writing has been archived at
\url{http://www.webcitation.org/6CTpDz4BX}}}
\begin{ruledtabular}
\begin{tabular}{crll}
 lattice & $z$ & $p_c$ (site) & $p_c$ (bond) \\
 \hline
  (8,3)-a & 3 & $0.577962(33)$ & $0.555700(22)$ \\
 (10,3)-a & 3 & $0.571404(40)$ & $0.551060(37)$ \\
 (10,3)-b & 3 & $0.565442(40)$ & $0.546694(33)$ \\
 %(12,3) & 3 & $0.56548(4)$ & $? \pm ?$ \\
 diamond & 4 & $0.4301(4)$\cite{VanDerMarck1998} & $0.3893(2)$\cite{VanDerMarck1998} \\
 simple cubic & 6 & $0.3116080(4)$\cite{Lorenz1998a} & $0.2488126(5)$\cite{Lorenz1998} \\
 bcc & 8 & $0.2459615(10)$\cite{Lorenz1998a} & $0.1802875(10)$\cite{Lorenz1998} \\
 fcc & 12 & $0.1992365(10)$\cite{Lorenz1998} & $0.1201635(10)$\cite{Lorenz1998a} \\
 hcp & 12 & $0.1992555(10)$\cite{Lorenz2000} & $0.1201640(10)$\cite{Lorenz2000} \\
\hline
\end{tabular}
\end{ruledtabular}
\end{table}

\section{Discussion}

The percolation thresholds identified for the 3-coordinated lattices
considered here are higher than typical values for 3-dimensional
lattices that have been studied previously.  This point is
illustrated in Table~\ref{tab:table1}, which shows percolation
thresholds for a variety of common 3-dimensional lattices, organized
by their coordination number $z$.  It is clear from the table that
$p_c$ increases as $z$ decreases.  This makes intuitive sense:  With
lower coordination numbers, it is easier to destroy a spanning
cluster by removing a few sites or bonds at key points, while in
lattices with higher coordination numbers there are more paths that
can be navigated to circumvent a missing site or bond.

\begin{figure}[htb]
 \includegraphics[scale=0.35]{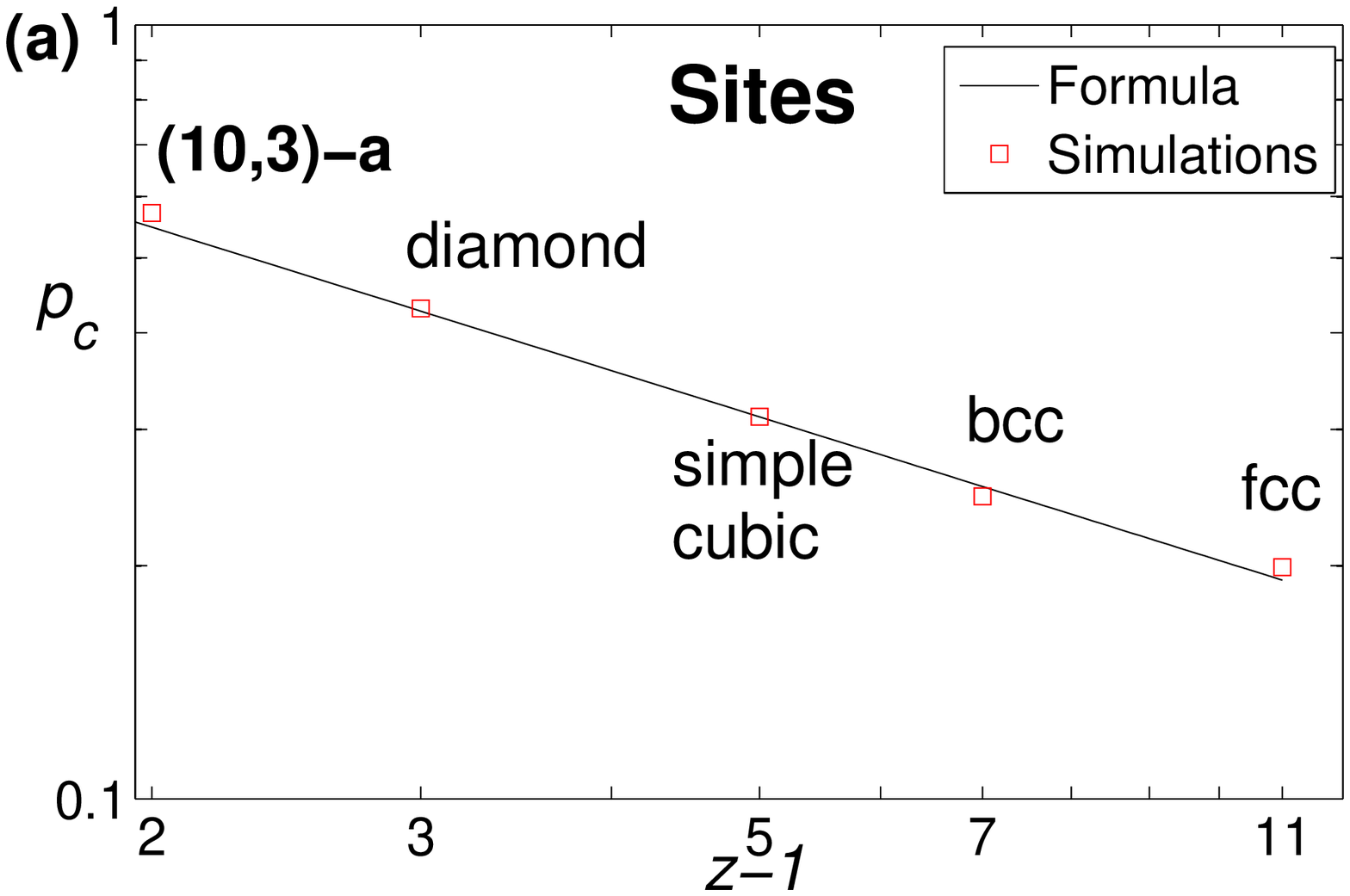}%
 \\
 \includegraphics[scale=0.35]{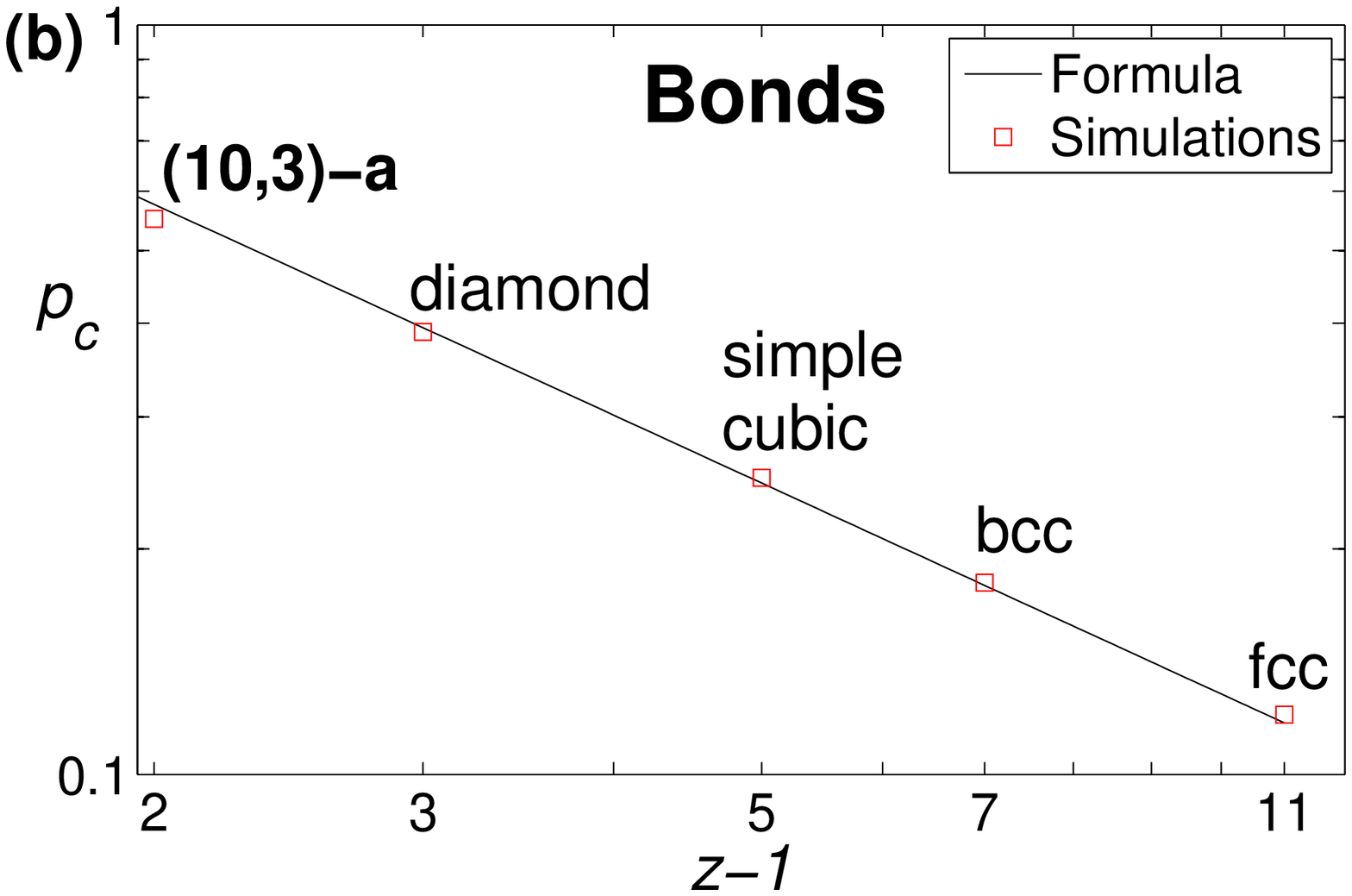}%
 \caption{\label{fig:scaling} A log-log plot of computationally-determined
 site (top) and bond (bottom) percolation thresholds \vs\
$(z-1)$ for different
 3D lattices, and a comparison plot of the formula of Galam and
 Mauger\cite{Galam1996}.  Simulation values for (10,3)-a are from \Fig{fig:103avsL}.
 Simulations values for the other lattices are from the references in Table~\ref{tab:table1}.}
\end{figure}

The information shown in Table~\ref{tab:table1} can be compared with
an analytical expression\cite{Galam1996} for the approximate
dependence of $p_c$ on spatial dimension $d$ and coordination number
$z$:
\begin{equation}
    p_c = p_0 \left((d-1)(z-1) \right)^{-a} d^{b}
\end{equation}
where $a= 0.6160$ for sites and 0.9346 bonds, $b=0$ for sites and
$b=a$ for bonds, and $p_0= 1.2868$ for sites and 0.7541 for bonds.
This comparison is done in \Fig{fig:scaling}.  For ease of
presentation we only show the (10,3)-a lattice on these plots, but
the percolation thresholds of the other 3-connected lattices differ
by only about $1\%$, and would overlap the (10,3)-a lattice if
included.  Crucially, the percolation thresholds of the 3-connected
lattices are very close to the theoretical plot.

We have not explored the entire family of 3-connected nets studied
by Wells.  Thus, it is an open question as to whether there are
simple 3-dimensional periodic lattices with higher percolation
thresholds.  We qualify the previous sentence with the word
``simple'' because it is intuitively obvious that one could often
trivially increase the percolation threshold of a crystalline
structure by inserting chains of 2-connected sites between sites
with higher coordination numbers.  The interesting question is
whether one can construct crystals with higher $p_c$ without making
use of 2-connected sites between higher-coordinated cites.

In conclusion, we have used Monte Carlo simulations to determine the
percolation thresholds of several interesting lattices that have
not, to the best of our knowledge, been studied previously.  We find
that these lattices have substantially higher percolation thresholds
than other 3-dimensional lattices, due to their low coordination
numbers.  The results for both bond and percolation thresholds are
very close to theoretical predictions that rely on coordination
number and dimensionality.

%\section{}

% If you have acknowledgments, this puts in the proper section head.
\begin{acknowledgments}
Jonathan Tran was supported by the Kellogg University Scholars
program. Shane Stahlheber was supported by the Microscopy Society of
America.  Alex Small was supported by a Teacher-Scholar Award from
California State Polytechnic University.
\end{acknowledgments}

% Create the reference section using BibTeX:
%\bibliography{biblio-perc}
%\bibliography{biblio}

\begin{thebibliography}{18}%
\makeatletter
\providecommand \@ifxundefined [1]{%
 \@ifx{#1\undefined}
}%
\providecommand \@ifnum [1]{%
 \ifnum #1\expandafter \@firstoftwo
 \else \expandafter \@secondoftwo
 \fi
}%
\providecommand \@ifx [1]{%
 \ifx #1\expandafter \@firstoftwo
 \else \expandafter \@secondoftwo
 \fi
}%
\providecommand \natexlab [1]{#1}%
\providecommand \enquote  [1]{``#1''}%
\providecommand \bibnamefont  [1]{#1}%
\providecommand \bibfnamefont [1]{#1}%
\providecommand \citenamefont [1]{#1}%
\providecommand \href@noop [0]{\@secondoftwo}%
\providecommand \href [0]{\begingroup \@sanitize@url \@href}%
\providecommand \@href[1]{\@@startlink{#1}\@@href}%
\providecommand \@@href[1]{\endgroup#1\@@endlink}%
\providecommand \@sanitize@url [0]{\catcode `\\12\catcode
`\$12\catcode
  `\&12\catcode `\#12\catcode `\^12\catcode `\_12\catcode `\%12\relax}%
\providecommand \@@startlink[1]{}%
\providecommand \@@endlink[0]{}%
\providecommand \url  [0]{\begingroup\@sanitize@url \@url }%
\providecommand \@url [1]{\endgroup\@href {#1}{\urlprefix }}%
\providecommand \urlprefix  [0]{URL }%
\providecommand \Eprint [0]{\href }%
\providecommand \doibase [0]{http://dx.doi.org/}%
\providecommand \selectlanguage [0]{\@gobble}%
\providecommand \bibinfo  [0]{\@secondoftwo}%
\providecommand \bibfield  [0]{\@secondoftwo}%
\providecommand \translation [1]{[#1]}%
\providecommand \BibitemOpen [0]{}%
\providecommand \bibitemStop [0]{}%
\providecommand \bibitemNoStop [0]{.\EOS\space}%
\providecommand \EOS [0]{\spacefactor3000\relax}%
\providecommand \BibitemShut  [1]{\csname bibitem#1\endcsname}%
\let\auto@bib@innerbib\@empty
%</preamble>
\bibitem [{\citenamefont {Stauffer}\ and\ \citenamefont
  {Aharony}(1994)}]{Stauffer1994}%
  \BibitemOpen
  \bibfield  {author} {\bibinfo {author} {\bibfnamefont {D.}~\bibnamefont
  {Stauffer}}\ and\ \bibinfo {author} {\bibfnamefont {A.}~\bibnamefont
  {Aharony}},\ }\href@noop {} {\emph {\bibinfo {title} {Introduction to
  Percolation Theory}}},\ \bibinfo {edition} {2nd}\ ed.\ (\bibinfo  {publisher}
  {CRC},\ \bibinfo {year} {1994})\ p.\ \bibinfo {pages} {192}\BibitemShut
  {NoStop}%
\bibitem [{\citenamefont {Sahimi}(1994)}]{Sahimi1994}%
  \BibitemOpen
  \bibfield  {author} {\bibinfo {author} {\bibfnamefont {M.}~\bibnamefont
  {Sahimi}},\ }\href@noop {} {\emph {\bibinfo {title} {Applications of
  percolation theory}}}\ (\bibinfo  {publisher} {CRC},\ \bibinfo {year}
  {1994})\BibitemShut {NoStop}%
\bibitem [{\citenamefont {Galam}\ and\ \citenamefont
  {Mauger}(1996)}]{Galam1996}%
  \BibitemOpen
  \bibfield  {author} {\bibinfo {author} {\bibfnamefont {S.}~\bibnamefont
  {Galam}}\ and\ \bibinfo {author} {\bibfnamefont {A.}~\bibnamefont {Mauger}},\
  }\href {\doibase 10.1103/PhysRevE.53.2177} {\bibfield  {journal} {\bibinfo
  {journal} {Phys. Rev. E}\ }\textbf {\bibinfo {volume} {53}},\ \bibinfo
  {pages} {2177} (\bibinfo {year} {1996})}\BibitemShut {NoStop}%
\bibitem [{\citenamefont {Silverman}\ and\ \citenamefont
  {Adler}(1990)}]{Silverman1990}%
  \BibitemOpen
  \bibfield  {author} {\bibinfo {author} {\bibfnamefont {A.}~\bibnamefont
  {Silverman}}\ and\ \bibinfo {author} {\bibfnamefont {J.}~\bibnamefont
  {Adler}},\ }\href {\doibase 10.1103/PhysRevB.42.1369} {\bibfield  {journal}
  {\bibinfo  {journal} {Physical Review B}\ }\textbf {\bibinfo {volume} {42}},\
  \bibinfo {pages} {1369} (\bibinfo {year} {1990})}\BibitemShut {NoStop}%
\bibitem [{\citenamefont {Vyssotsky}\ \emph {et~al.}(1961)\citenamefont
  {Vyssotsky}, \citenamefont {Gordon}, \citenamefont {Frisch},\ and\
  \citenamefont {Hammersley}}]{Vyssotsky1961}%
  \BibitemOpen
  \bibfield  {author} {\bibinfo {author} {\bibfnamefont {V.}~\bibnamefont
  {Vyssotsky}}, \bibinfo {author} {\bibfnamefont {S.}~\bibnamefont {Gordon}},
  \bibinfo {author} {\bibfnamefont {H.}~\bibnamefont {Frisch}}, \ and\ \bibinfo
  {author} {\bibfnamefont {J.}~\bibnamefont {Hammersley}},\ }\href {\doibase
  10.1103/PhysRev.123.1566} {\bibfield  {journal} {\bibinfo  {journal}
  {Physical Review}\ }\textbf {\bibinfo {volume} {123}},\ \bibinfo {pages}
  {1566} (\bibinfo {year} {1961})}\BibitemShut {NoStop}%
\bibitem [{Note1()}]{Note1}%
  \BibitemOpen
  \bibinfo {note} {In order to have an average coordination number less than 3,
  one would need at least some sites with $z=2$, but a site with $z=2$ is
  equivalent to a single bond between two other sites. A physical analogy would
  be the role of an oxygen atom in an SiO$_2$ crystal. So, any lattice with
  average $z < 3$ must be equivalent to taking a lattice with $z\geq 3$ and
  placing sites with $z=2$ along some of the bonds.}\BibitemShut {Stop}%
\bibitem [{\citenamefont {Wells}(1977)}]{Wells1977}%
  \BibitemOpen
  \bibfield  {author} {\bibinfo {author} {\bibfnamefont {A.}~\bibnamefont
  {Wells}},\ }\href@noop {} {\emph {\bibinfo {title} {Three dimensional nets
  and polyhedra}}}\ (\bibinfo  {publisher} {Wiley},\ \bibinfo {year}
  {1977})\BibitemShut {NoStop}%
\bibitem [{\citenamefont {Paterson}\ \emph {et~al.}(2002)\citenamefont
  {Paterson}, \citenamefont {Sheppard},\ and\ \citenamefont
  {Knackstedt}}]{Paterson2002}%
  \BibitemOpen
  \bibfield  {author} {\bibinfo {author} {\bibfnamefont {L.}~\bibnamefont
  {Paterson}}, \bibinfo {author} {\bibfnamefont {A.~P.}\ \bibnamefont
  {Sheppard}}, \ and\ \bibinfo {author} {\bibfnamefont {M.~A.}\ \bibnamefont
  {Knackstedt}},\ }\href {\doibase 10.1103/PhysRevE.66.056122} {\bibfield
  {journal} {\bibinfo  {journal} {Phys. Rev. E}\ }\textbf {\bibinfo {volume}
  {66}},\ \bibinfo {pages} {056122} (\bibinfo {year} {2002})}\BibitemShut
  {NoStop}%
\bibitem [{\citenamefont {Sunada}(2008)}]{Sunada2008}%
  \BibitemOpen
  \bibfield  {author} {\bibinfo {author} {\bibfnamefont {T.}~\bibnamefont
  {Sunada}},\ }\href@noop {} {\bibfield  {journal} {\bibinfo  {journal}
  {Notices of the AMS}\ }\textbf {\bibinfo {volume} {55}},\ \bibinfo {pages}
  {208} (\bibinfo {year} {2008})}\BibitemShut {NoStop}%
\bibitem [{\citenamefont {Bates}(2005)}]{Bates2005}%
  \BibitemOpen
  \bibfield  {author} {\bibinfo {author} {\bibfnamefont {F.}~\bibnamefont
  {Bates}},\ }\href@noop {} {\bibfield  {journal} {\bibinfo  {journal} {Mater
  Res Bull}\ }\textbf {\bibinfo {volume} {30}},\ \bibinfo {pages} {525}
  (\bibinfo {year} {2005})}\BibitemShut {NoStop}%
\bibitem [{\citenamefont {Itoh}\ \emph {et~al.}(2009)\citenamefont {Itoh},
  \citenamefont {Kotani}, \citenamefont {Naito}, \citenamefont {Sunada},
  \citenamefont {Kawazoe},\ and\ \citenamefont {Adschiri}}]{Itoh2009}%
  \BibitemOpen
  \bibfield  {author} {\bibinfo {author} {\bibfnamefont {M.}~\bibnamefont
  {Itoh}}, \bibinfo {author} {\bibfnamefont {M.}~\bibnamefont {Kotani}},
  \bibinfo {author} {\bibfnamefont {H.}~\bibnamefont {Naito}}, \bibinfo
  {author} {\bibfnamefont {T.}~\bibnamefont {Sunada}}, \bibinfo {author}
  {\bibfnamefont {Y.}~\bibnamefont {Kawazoe}}, \ and\ \bibinfo {author}
  {\bibfnamefont {T.}~\bibnamefont {Adschiri}},\ }\href@noop {} {\bibfield
  {journal} {\bibinfo  {journal} {Physical Review Letters}\ }\textbf {\bibinfo
  {volume} {102}},\ \bibinfo {pages} {055703} (\bibinfo {year} {2009})},\
  \bibinfo {note} {pRL}\BibitemShut {NoStop}%
\bibitem [{\citenamefont {Newman}\ and\ \citenamefont
  {Ziff}(2001)}]{Newman2001}%
  \BibitemOpen
  \bibfield  {author} {\bibinfo {author} {\bibfnamefont {M.~E.~J.}\
  \bibnamefont {Newman}}\ and\ \bibinfo {author} {\bibfnamefont {R.~M.}\
  \bibnamefont {Ziff}},\ }\href@noop {} {\bibfield  {journal} {\bibinfo
  {journal} {Physical Review E}\ }\textbf {\bibinfo {volume} {64}},\ \bibinfo
  {pages} {016706} (\bibinfo {year} {2001})},\ \bibinfo {note}
  {pRE}\BibitemShut {NoStop}%
\bibitem [{\citenamefont {Ziff}\ and\ \citenamefont {Newman}(2002)}]{Ziff2002}%
  \BibitemOpen
  \bibfield  {author} {\bibinfo {author} {\bibfnamefont {R.~M.}\ \bibnamefont
  {Ziff}}\ and\ \bibinfo {author} {\bibfnamefont {M.~E.~J.}\ \bibnamefont
  {Newman}},\ }\href@noop {} {\bibfield  {journal} {\bibinfo  {journal}
  {Physical Review E}\ }\textbf {\bibinfo {volume} {66}},\ \bibinfo {pages}
  {016129} (\bibinfo {year} {2002})},\ \bibinfo {note} {pRE}\BibitemShut
  {NoStop}%
\bibitem [{\citenamefont {Martins}\ and\ \citenamefont
  {Plascak}(2003)}]{Martins2003}%
  \BibitemOpen
  \bibfield  {author} {\bibinfo {author} {\bibfnamefont {P.~H.~L.}\
  \bibnamefont {Martins}}\ and\ \bibinfo {author} {\bibfnamefont {J.~A.}\
  \bibnamefont {Plascak}},\ }\href@noop {} {\bibfield  {journal} {\bibinfo
  {journal} {Physical Review E}\ }\textbf {\bibinfo {volume} {67}},\ \bibinfo
  {pages} {046119} (\bibinfo {year} {2003})},\ \bibinfo {note}
  {pRE}\BibitemShut {NoStop}%
\bibitem [{\citenamefont {Van Der~Marck}(1998)}]{VanDerMarck1998}%
  \BibitemOpen
  \bibfield  {author} {\bibinfo {author} {\bibfnamefont {S.}~\bibnamefont {Van
  Der~Marck}},\ }\href@noop {} {\bibfield  {journal} {\bibinfo  {journal}
  {International Journal of Modern Physics C}\ }\textbf {\bibinfo {volume}
  {9}},\ \bibinfo {pages} {529} (\bibinfo {year} {1998})}\BibitemShut {NoStop}%
\bibitem [{\citenamefont {Lorenz}\ and\ \citenamefont
  {Ziff}(1998{\natexlab{a}})}]{Lorenz1998a}%
  \BibitemOpen
  \bibfield  {author} {\bibinfo {author} {\bibfnamefont {C.~D.}\ \bibnamefont
  {Lorenz}}\ and\ \bibinfo {author} {\bibfnamefont {R.~M.}\ \bibnamefont
  {Ziff}},\ }\href {\doibase 10.1088/0305-4470/31/40/009} {\bibfield  {journal}
  {\bibinfo  {journal} {Journal of Physics A: Mathematical and General}\
  }\textbf {\bibinfo {volume} {31}},\ \bibinfo {pages} {8147} (\bibinfo {year}
  {1998}{\natexlab{a}})}\BibitemShut {NoStop}%
\bibitem [{\citenamefont {Lorenz}\ and\ \citenamefont
  {Ziff}(1998{\natexlab{b}})}]{Lorenz1998}%
  \BibitemOpen
  \bibfield  {author} {\bibinfo {author} {\bibfnamefont {C.}~\bibnamefont
  {Lorenz}}\ and\ \bibinfo {author} {\bibfnamefont {R.}~\bibnamefont {Ziff}},\
  }\href {\doibase 10.1103/PhysRevE.57.230} {\bibfield  {journal} {\bibinfo
  {journal} {Physical Review E}\ }\textbf {\bibinfo {volume} {57}},\ \bibinfo
  {pages} {230} (\bibinfo {year} {1998}{\natexlab{b}})}\BibitemShut {NoStop}%
\bibitem [{\citenamefont {Lorenz}\ \emph {et~al.}(2000)\citenamefont {Lorenz},
  \citenamefont {May},\ and\ \citenamefont {Ziff}}]{Lorenz2000}%
  \BibitemOpen
  \bibfield  {author} {\bibinfo {author} {\bibfnamefont {C.~D.}\ \bibnamefont
  {Lorenz}}, \bibinfo {author} {\bibfnamefont {R.}~\bibnamefont {May}}, \ and\
  \bibinfo {author} {\bibfnamefont {R.~M.}\ \bibnamefont {Ziff}},\ }\href@noop
  {} {\bibfield  {journal} {\bibinfo  {journal} {Journal of Statistical
  Physics}\ }\textbf {\bibinfo {volume} {98}},\ \bibinfo {pages} {961}
  (\bibinfo {year} {2000})}\BibitemShut {NoStop}%
\end{thebibliography}
%

\end{document}